\documentclass[superscriptaddress,aps,prd,showpacs,nofootinbib,floatfix,twocolumn]{revtex4}
\usepackage{amsmath}
\usepackage{amssymb}
\usepackage{graphicx}
\usepackage[english]{babel}
\usepackage{multirow}
\usepackage{xspace}
\usepackage[sc]{mathpazo}

\newcommand\Xmum{\ensuremath{X_\text{max}^\mu}\xspace}
\newcommand\Xmumm{\ensuremath{X_\text{max}^\mu}}
\newcommand\Xm{\ensuremath{X_\text{max}}\xspace}
\newcommand\Xmm{\ensuremath{X_\text{max}}}
\newcommand\Xmu{\ensuremath{X^\mu}\xspace}
\newcommand\Xmus{\ensuremath{X^\mu}}
\newcommand\QII{\textsc{QGSJetII-04}\xspace}
\newcommand\Epos{\textsc{Epos-LHC}\xspace}
\newcommand\Fluka{\textsc{Fluka} 2011.2b.4\xspace}

\begin{document}

\title{Muons in air showers at the Pierre Auger Observatory:
  Measurement of atmospheric production depth}


\author{A.~Aab}
\affiliation{Universit\"{a}t Siegen, Siegen, 
Germany}
\author{P.~Abreu}
\affiliation{Laborat\'{o}rio de Instrumenta\c{c}\~{a}o e F\'{\i}sica 
Experimental de Part\'{\i}culas - LIP and  Instituto Superior 
T\'{e}cnico - IST, Universidade de Lisboa - UL, 
Portugal}
\author{M.~Aglietta}
\affiliation{Osservatorio Astrofisico di Torino  (INAF), 
Universit\`{a} di Torino and Sezione INFN, Torino, 
Italy}
\author{M.~Ahlers}
\affiliation{University of Wisconsin, Madison, WI, 
USA}
\author{E.J.~Ahn}
\affiliation{Fermilab, Batavia, IL, 
USA}
\author{I.~Al Samarai}
\affiliation{Institut de Physique Nucl\'{e}aire d'Orsay (IPNO), 
Universit\'{e} Paris 11, CNRS-IN2P3, Orsay, 
France}
\author{I.F.M.~Albuquerque}
\affiliation{Universidade de S\~{a}o Paulo, Instituto de F\'{\i}sica, 
S\~{a}o Paulo, SP, 
Brazil}
\author{I.~Allekotte}
\affiliation{Centro At\'{o}mico Bariloche and Instituto Balseiro 
(CNEA-UNCuyo-CONICET), San Carlos de Bariloche, 
Argentina}
\author{J.~Allen}
\affiliation{New York University, New York, NY, 
USA}
\author{P.~Allison}
\affiliation{Ohio State University, Columbus, OH, 
USA}
% ...
\collaboration{The Pierre Auger Collaboration}
\email{{\tt auger\_spokespersons@fnal.gov}}
\noaffiliation


\begin{abstract}
The surface detector array of the Pierre Auger Observatory provides
information about the longitudinal development of the muonic component
of extensive air showers. Using the timing information from the flash
analog-to-digital converter traces of surface detectors far from the
shower core, it is possible to reconstruct a muon production depth
distribution. We characterize the goodness of this reconstruction for
zenith angles around 60$^{\circ}$ and different energies of the primary particle. From
these distributions we define \Xmum as the depth along the shower axis
where the production of muons reaches maximum. We explore the
potentiality of \Xmum as a useful observable to infer the mass
composition of ultrahigh-energy cosmic rays. Likewise, we assess its
ability to constrain hadronic interaction models.  
\end{abstract}

\pacs{96.50.sd, 98.70.Sa, 13.85.Tp}

\maketitle
\section{Introduction}
Cosmic rays with energies on the joule scale are one of the most
intriguing subjects of fundamental physics in the 21st
century. Although the first indications of their existence were
obtained more than fifty years ago~\cite{Linsley:1961kt}, many of
their properties remain mysterious~\cite{LetessierSelvon:2011dy}. To
answer the question of the origin of ultrahigh-energy cosmic rays
(UHECRs) requires three experimental feats: finding the mass of the
primary particles, measuring the energy spectrum, and measuring the
distribution of arrival directions. The energy spectrum is the best
known of the three and its main features are well
established~\cite{Abbasi:2009ix,Abraham:2010mj}.  

The situation regarding the mass composition of UHECRs is not
settled. Since we have no direct access to the primary particle,
identifying its mass is a rather difficult endeavor. Mass may be
inferred from comparisons of the measured observables with Monte Carlo
simulations. These simulations rely on extrapolations of
accelerator data gathered at energies which are orders of magnitude
below those of UHECRs. Therefore, simulations constitute the most
prominent source of systematic uncertainties.  

One possible way to determine the mass is to study the longitudinal
development of the electromagnetic component of a shower. The depth of
the shower maximum \Xm is sensitive to the nature of the primary
particles~\cite{Baltrusaitis,Heitler,Abraham:2010yv}. It also helps to
provide insight on whether new physics phenomena take place at these
extreme energies. The measurements of \Xm performed by the Pierre
Auger and HiRes/TA collaborations cannot be compared directly
because of the different detector acceptances and the different \Xm
analysis approaches. However, when interpreted in terms of
composition, the Pierre Auger Collaboration claims evidence of a light
composition at energies around 3\,EeV and a gradual increase of the
average mass of cosmic rays towards higher energies~\cite{massJCAP},
while the HiRes/TA collaborations cannot currently discriminate
between a proton-dominated composition and a changing composition (such
as suggested by Auger), due to statistical limitations~\cite{cern12}. 

 \Xm measurements suffer from low statistics due to the small duty
 cycle of fluorescence detectors (less than 15\%) and the stringent
 cuts imposed to avoid a biased data sample. Therefore, to gain
 additional insight on questions like mass composition or whether new
 hadronic interactions are taking place, we need independent
 measurements with larger statistics, a different set of systematic
 uncertainties, and the possibility of reaching higher energies. To
 infer the mass of the primary,  we can study the longitudinal
 development of the muonic component of extensive air showers
 (EASs)~\cite{Diego:2013}, thus taking advantage of the large
 statistical sample provided by the Auger surface detector (SD)
 array. The SD array delivers this information through the timing records
 associated with the muons that reach the ground. The arrival times of
 the muons allow the reconstruction of their geometrical production
 heights along the shower axis. It is thus possible to reconstruct a
 distribution of muon production depths (MPD). Since muons come from
 the decay of pions and kaons, the shape of the MPD distribution
 contains information about the evolution of the hadronic
 cascade. This information renders the study of MPD interesting for
 two reasons: on the one hand, we know that different primaries have
 distinct hadronic properties that translate into variations of their
 respective longitudinal profiles. Therefore, it is natural to think
 that the MPD distribution must be sensitive to the mass of primary
 particles. On the other hand, the MPD distribution might be an
 optimal tool to study hadronic interactions at ultrahigh-energies
 since the longitudinal development is dependent on the hadronic
 interaction properties. MPD distributions can help in understanding
 whether a departure from standard physics is the source of the
 substantial differences observed between data and the current models
 used for shower simulations~\cite{AugerMuons}. In this work we
 explore the possibility of using MPD distributions as an experimental
 observable sensitive to the mass of the primary cosmic rays and able
 to constrain high-energy interaction models. The Auger Collaboration
 is currently evaluating other methods based on surface detector data
 ~\cite{Diego:2011} that could add valuable information to the set of
 parameters sensitive to mass composition.
\section{The Pierre Auger Observatory}
The Pierre Auger Observatory is located in the Province of Mendoza,
Argentina (35.1$^\circ$--35.5$^\circ$\,S, 69.0$^\circ$--69.6$^\circ$\,W,
about 1400\,m above see level).  Two detection methods are used to obtain
information about EASs, and hence information on the primary cosmic rays
that create them. The SD array is comprised of 1660 cylindrical
water-Cherenkov detectors arranged on a triangular grid, with 1500\,m
spacing, that covers an area of over 3000\,km$^2$. Each detector has a
10\,m$^2$ surface area and 1.2\,m water depth, the water volume being
viewed by three 9\,inch photomultiplier tubes
(PMTs)~\cite{Abraham:2004dt,Allekotte:2007sf}. PMT signals are
digitized using 40\,MHz, 10 bit flash analog-to-digital converters
(FADCs). The detectors respond to the muons, photons, and electrons of
air showers and are calibrated in units of the signal produced by a
muon traversing the water vertically, known as a {\it vertical
  equivalent muon} or VEM~\cite{Bertou:2005ze}. The fluorescence
detector (FD) consists of 27 optical telescopes overlooking the SD
array~\cite{Abraham:2009pm,fd2}. On clear moonless nights these are
used to observe the longitudinal development of showers by detecting
the fluorescence and Cherenkov light produced in the atmosphere by
charged particles along the shower trajectory. In the context of
primary mass studies, hybrid events have been used to provide a direct
measurement of \Xmm~\cite{Abraham:2010yv}.

However, the  bulk of events  collected by the observatory have
information only from the surface array, making SD observables, such
as the one described in this work, very valuable for composition
analysis at the highest  energies.  Only brief details of the
reconstruction methods are given here. More extended descriptions of
detectors and of reconstruction procedures can be found
in Refs.~\cite{Abraham:2004dt,Abraham:2010mj,Abraham:2010yv}. 
The trigger requirement for the surface array to form an event is
based on a threefold coincidence, satisfied when a triangle of
neighboring stations is triggered locally~\cite{Abraham:2010zz}. For
the present analysis, we use events that satisfy a fiducial cut to
ensure adequate containment inside the array. For events whose
reconstructed energy is above 3\,EeV, the efficiency of detection is
100\%. For SD data, the arrival directions are obtained from the times
at which the shower front passes through the triggered detectors, this
time being measured using GPS information. The angular resolution,
defined as the angular radius around the true cosmic-ray direction
that would contain 68\% of the reconstructed shower directions, is
0.8$^\circ$ for energies above 3\,EeV~\cite{Bonifazi:2009ma}. The
estimator of the primary energy of events recorded by the SD is the
reconstructed signal at 1000\,m from the shower core, $S$(1000). The
conversion from this estimator to energy is derived experimentally
through the use of a subset of showers that trigger the FD and the SD
simultaneously (hybrid events). The energy resolution above 10\,EeV is
about 12\%. The absolute energy scale, determined by the FD, has a systematic
uncertainty of 14\%~\cite{ValerioICRC13}. 
\section{Reconstruction of the Muon Production Depth Distribution}
When an EAS develops in the atmosphere, the
transverse momentum of secondary 
particles makes them deviate from the shower axis on their way to 
the ground. Unlike the electromagnetic component of the shower, 
muon trajectories can be taken as straight lines, 
due to the lesser importance of bremsstrahlung and
 multiple scattering effects. This fact confers the muons a
 distinctive
 attribute: they retain a memory of
 their production points. The muon component reaching the ground has a time
 structure caused by the convolution of production spectra, energy
 loss, and decay probability during propagation. Thanks to a set
 of simple assumptions~\cite{LCThesis}, these arrival times can be
 used to obtain the distribution of muon production distances along
 the shower axis. Since muons are the products of pion and kaon
 decays, the distribution of muon production distances provides
 information about the longitudinal development of the
 hadronic component of the EAS~\cite{kascade}.  This information is
 complementary to that obtained from the electromagnetic component
 through the detection of atmospheric fluorescence light.  
\begin{figure}
\centering
\includegraphics[width=\columnwidth]{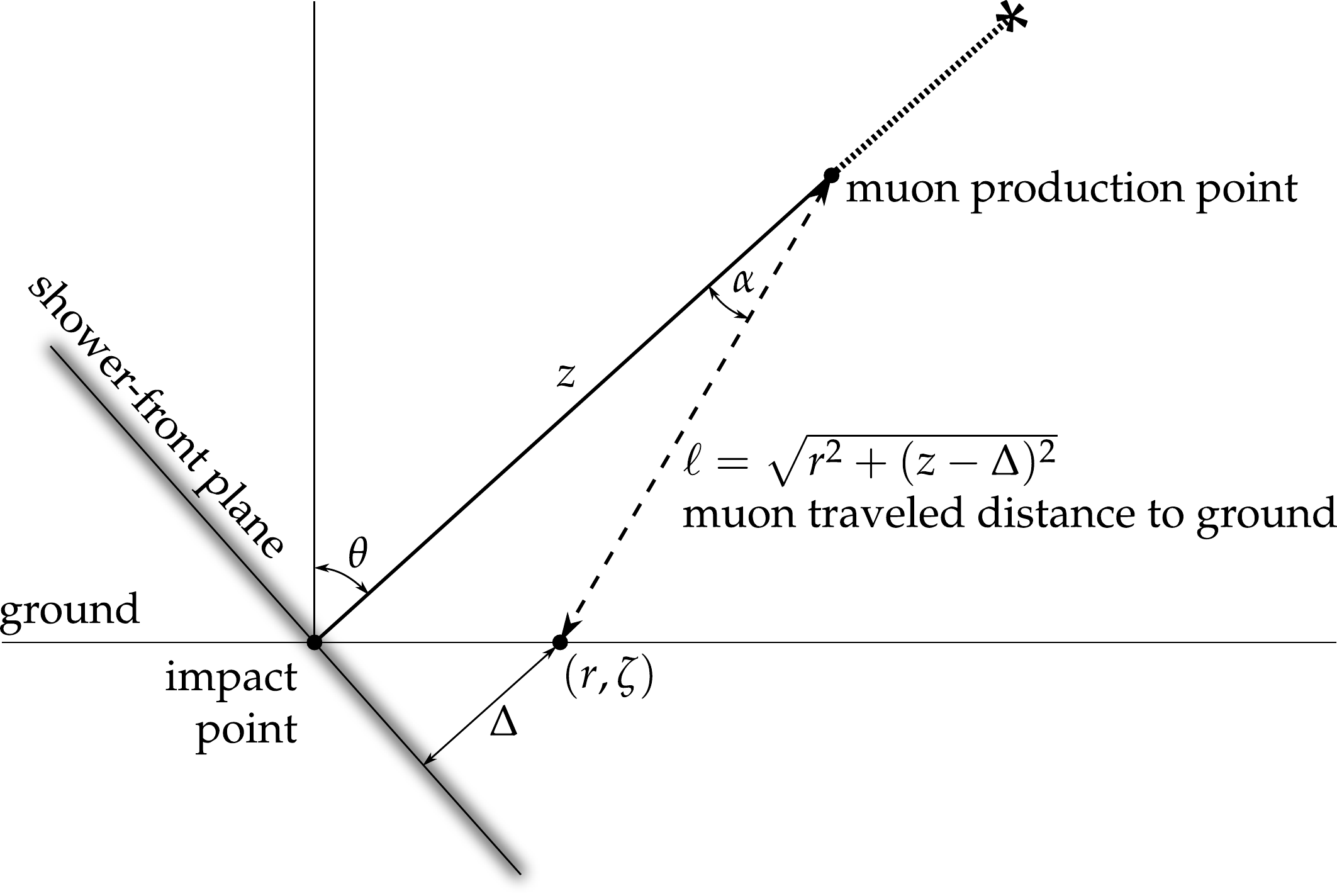}
\caption{Geometry used to obtain the muon traveled
distance and the time delay.
\label{fig:_geometry}}
\end{figure}

The basis of our measurement is a theoretical framework originally
developed  in Refs.~\cite{Cazon:2003ar,Cazon:2004zx} and updated
in Ref.~\cite{Cazon:2012mt} to model the muon distributions in
EAS. Here we summarize its main aspects.  
As a first approximation, we
assume that muons travel in straight lines at the speed of light $c$
and that they are produced in the shower axis. 
This is outlined in Figure~\ref{fig:_geometry},
where muons are produced at the position $z$ along the shower axis
and, after traveling a distance $\ell$, they reach the ground at the point defined by
$(r,\zeta)$. $r$ and $\zeta$ are measured in the shower reference
frame and represent the distance and the azimuthal position of the
point at the ground, respectively. $\Delta$ is the distance from the
ground impact point to the shower plane. Referencing the muon time of flight
to the arrival time of the shower-front plane for each position
($r,\zeta$), we obtain what we define as the {\it geometric delay},
$t_\text{g}$. It represents the delay of muons due to the deviation of
their trajectories with respect to the direction of the shower
axis. Given $t_\text{g}$ it is possible to derive the production
distance $z$ of muons for each position ($r,\zeta$) at the ground. 

The geometric delay is not the only source contributing to 
the measured muon delay $t$. The average energy of muons at
production ($v_\mu<c$) and their energy loss, mainly because of
inelastic collisions with atomic electrons in the air, 
cause a {\it kinematic delay} $t_\varepsilon$, with respect to a
particle traveling at the speed of light. To compute it, we need an
estimation of the energy carried by each single muon. The Auger SD
array does not allow for such a measurement: therefore
we must use for this correction a mean kinematic time value
$\langle t_\varepsilon \rangle$ as an
approximation~\cite{Cazon:2004zx}. An additional source of delay is given
by the deflection of muons 
due to their elastic scattering off
nuclei. Furthermore, the geomagnetic field affects the
trajectory of the muons, delaying their arrival times even more. The longer the
path of the muon, the larger is the effect; hence, it is especially important for
very inclined events.  

\begin{figure}
\centering
\includegraphics[width=\columnwidth]{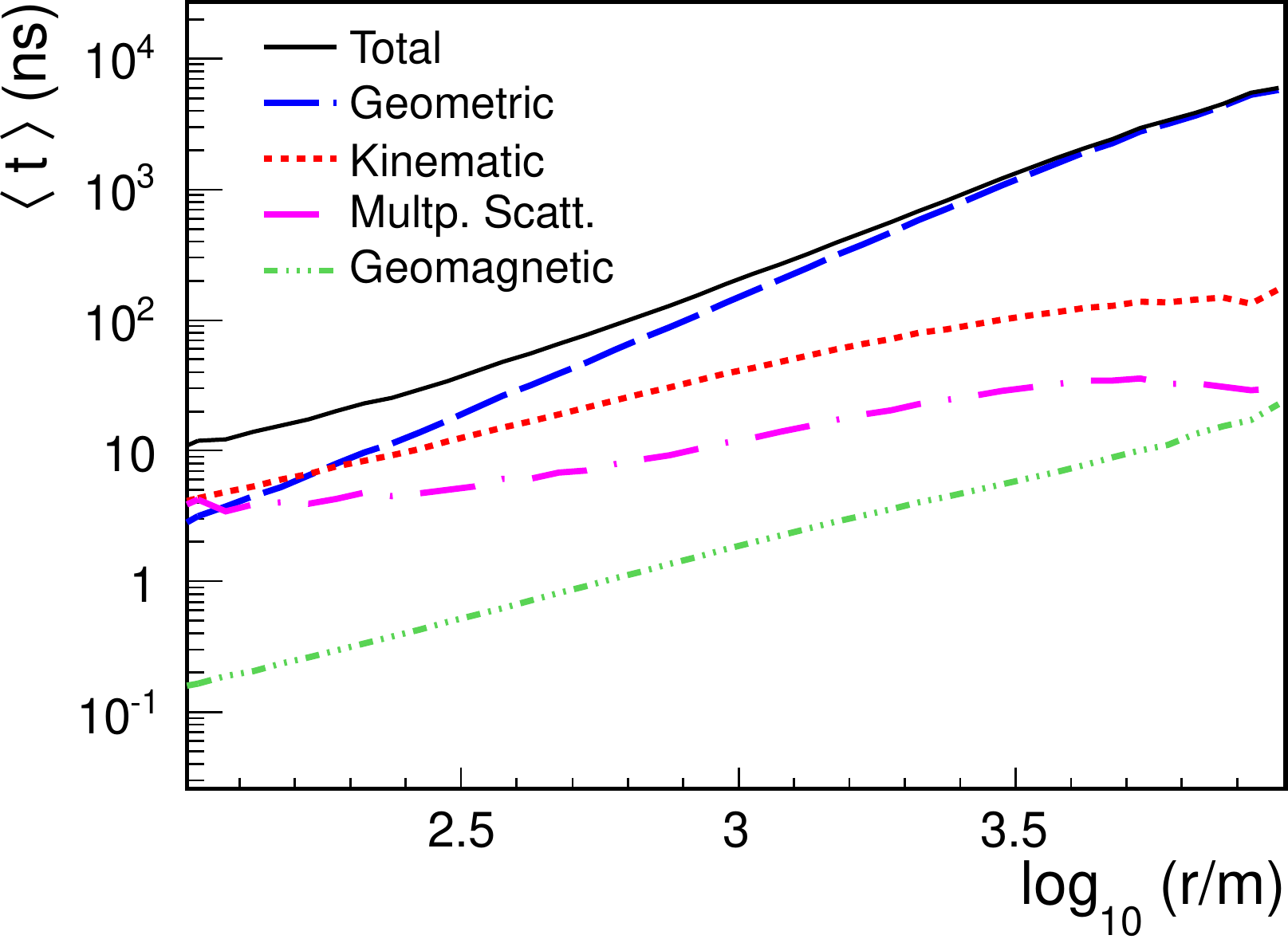}
\caption{Average time delay of muons with a breakdown of the different
  contributions. Those muons are produced in a proton-initiated
  shower with a zenith angle of 60$^{\circ}$ and primary energy of $E$
  = 10\,EeV~\cite{Cazon:2012mt}.  
\label{fig:Delays}}
\end{figure}

To demonstrate the importance of the different contributions to the total
delay, Figure~\ref{fig:Delays} presents, for events at a 60$^{\circ}$
zenith angle, the average value of each delay as a function of the distance to
the shower core. All contributing effects show a clear dependence with $r$.  This
behavior is similar for events with different zenith angles. 
The geometric delay dominates at large distances. The contribution of
the kinematic effect is larger near the core. In principle, one may think
that the kinematic delay decreases closer to the core because muons
are more energetic on average. However, in this region the spread in
energy is larger~\cite{Cazon:2012mt} and the mean time delay is dominated
by low-energy muons. For events at $\sim$60$^{\circ}$, at distances $r
> 1000$\,m, the kinematic delay typically amounts to less than 30\% of
the total delay, while the rest of the contributions are of the order of a few percent
(see Figure~\ref{fig:Delays}). 

Since muons are not produced in the shower axis, we must apply a
correction due to the path traveled by the parent mesons. Assuming
that muons are collinear with the trajectory followed by the parent pion, the
muon paths start deeper in the atmosphere by an amount which is simply
the decay length of the pion: $z_\pi =
c\tau_\pi E_\pi/(m_\pi c^2) cos\alpha$. The pion energy dependence of
this correction has been taken from Ref.~\cite{Cazon:2004zx}. The distance
$z_\pi$ introduces an average time delay of
{$\sim$}3\,ns~\cite{Cazon:2012mt} (this correction amounts to
{$\sim$}1\% of the total delay; see Figure~\ref{fig:Delays}). 

All in all, the muon production point along the shower axis $z$ can be
inferred by the expression

\begin{equation}
z \simeq \frac{1}{2} \left( \frac{r^2}{c(t-\langle t_{\varepsilon} \rangle)}
-c(t-\langle t_{\varepsilon} \rangle)\right)+\Delta - \langle z_\pi \rangle,
\label{eq:zcorr_t}
\end{equation}
where the geometric delay $t_\text{g}$ has been approximated by $t_\text{g}\simeq t-\langle t_{\varepsilon} \rangle$.

For each point at the ground, Eq. \eqref{eq:zcorr_t} gives a mapping between the production distance $z$ and the arrival time $t$ of muons. The production distance can be easily related to the production depth \Xmu (total amount of traversed matter) using

\begin{equation}
X^{\mu} = \int_z^{\infty} \rho(z') \,\text{d}z',
\label{eq:ztoX}
\end{equation}
where $\rho$ stands for the atmospheric density. The set of production
depths forms the MPD distribution that describes the longitudinal
development of the muons generated in an air shower that reach the
ground. 
\section{Features of the muon profiles}
\label{s:rec}
The MPD is reconstructed from the FADC signals obtained with the water-Cherenkov detectors. The finite area of the detectors induces fluctuations due to different muon samples being collected. In addition, the shape of the MPD distribution observed from different positions at the ground varies because of differences in the probability of in-flight decay and because muons are not produced isotropically from the shower axis.
It is an integration over $r$ which enables estimation of the
$\text{d}N_{\mu}/\text{d}X$ distribution or MPD distribution (where
$N_{\mu}$ refers to the number of produced muons). However, for
discrete detector arrays, measurements at just a handful of $r$ values
are available and are limited to the small number of muons due to the
finite collection surface (10\,m$^{2}$ cross section for vertical incidence).

\begin{figure}
\centering
\includegraphics[width=\columnwidth,height=3in]{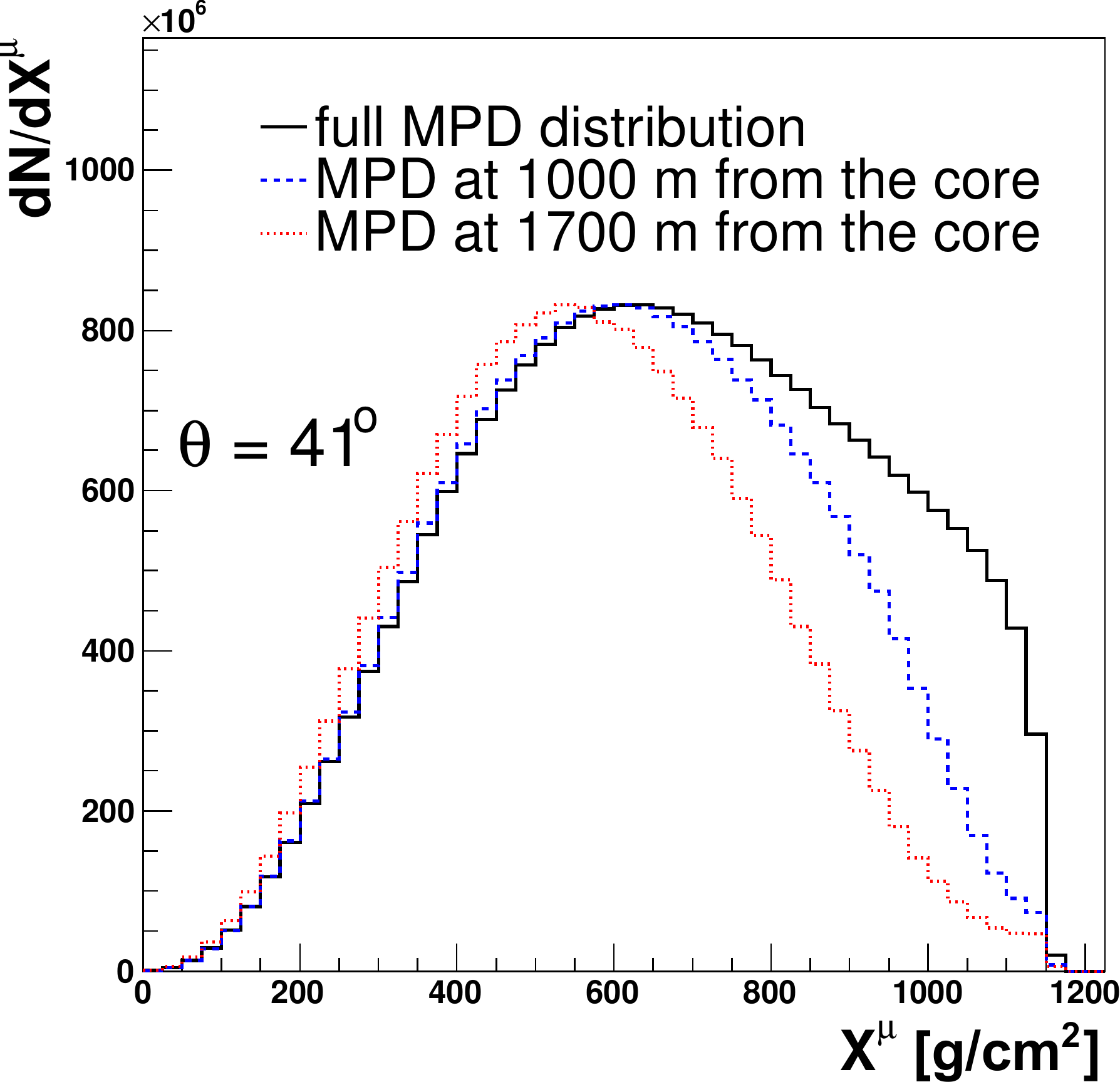}
\hfil
\includegraphics[width=\columnwidth,height=3in]{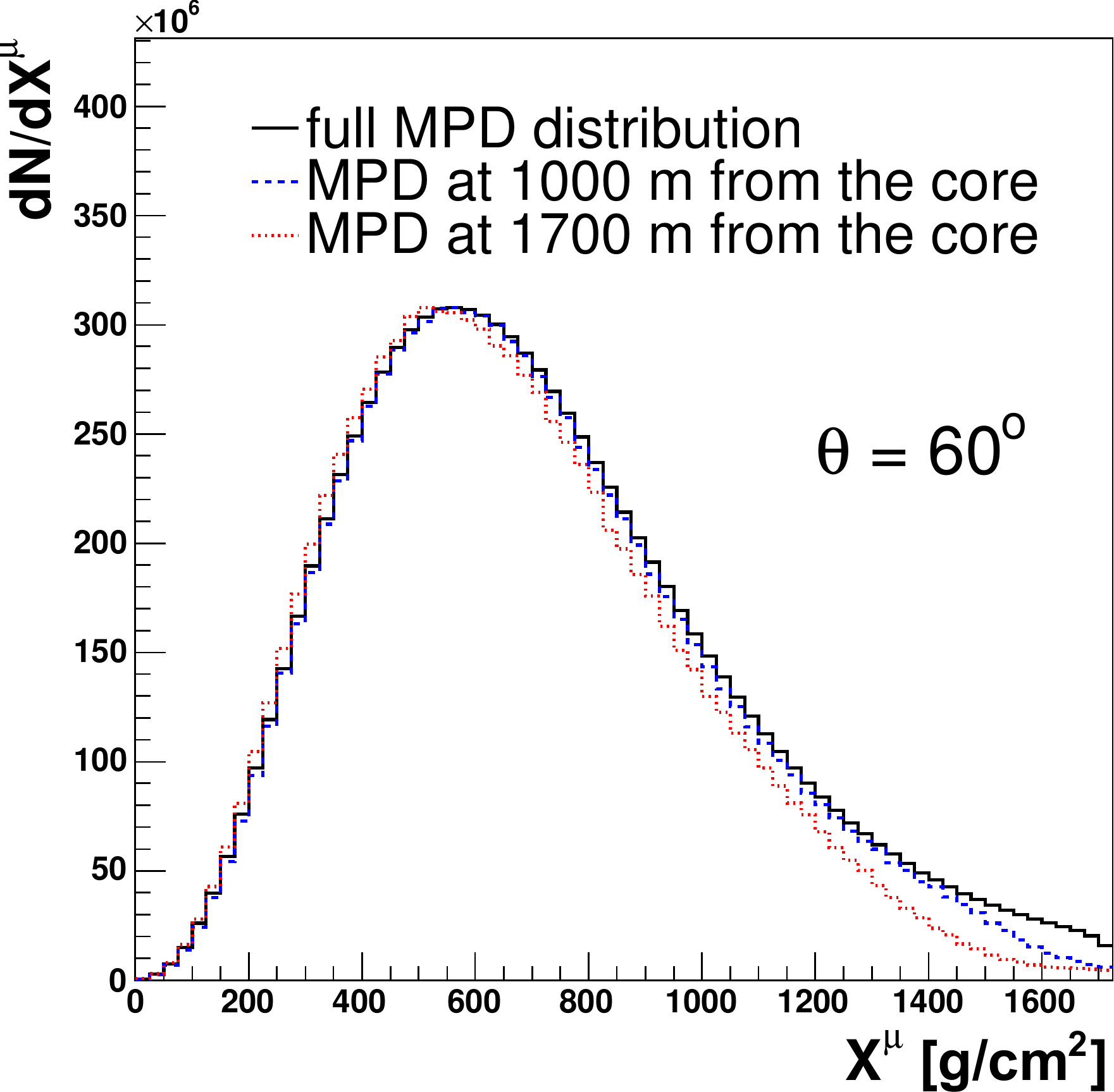}

\caption{MPD distributions produced by an iron shower with energy 10$^{19.5}$\,eV impinging with two different zenith angles: 41$^{\circ}$ (top) and 60$^{\circ}$ (bottom). We use \Epos~\cite{Epos} to model high-energy hadronic interactions. The shape of the MPD distribution shows a dependence with the distance to the shower core. This dependence is more pronounced for events with smaller inclinations. The histograms are normalized to have the same maximal height.
\label{fig:MPDvsr}}
\end{figure}
Another important feature of the MPD distribution observed at the
ground is its dependence on the zenith angle. There are two reasons
for this. The first one is due to the fact that inclined events mostly
evolve in a less dense atmosphere than more vertical ones. This makes
pions reach their critical energies ($\epsilon ^{\pi}_\text{c}$)
earlier, resulting in a shallower development of the shower. While the
shapes of the muon distributions at production are almost unaffected
by this difference in reaching $\epsilon ^{\pi}_\text{c}$
~\cite{Cazon:2012mt}, their production depths are shifted. For proton
showers at 10$^{19}$\,eV, the difference between the distribution of
maxima for vertical and  60$^{\circ}$ events is approximately
$20\,{\rm g}/{\rm cm}^2$. The second and main reason for the zenith
angle dependence of the MPD distribution is a consequence of the muon decay probability. This effect influences not only the location of the maximum but also the shape of the observed MPD distribution. Figure~\ref{fig:MPDvsr} demonstrates this dependence for MPDs extracted from simulations at different
zenith angles and at different distances $r$ from the core. For
zenith angles about 40$^{\circ}$ and lower, the shape of the MPD and the
position of its maximum are a function of $r$. However, at
zenith angles around 60$^{\circ}$ and above, the differences between the MPD distributions reconstructed 
at different distances to the core are small. At such angles, large $z$ values dominate, diminishing the dependence of the traveled muon distance $\ell$ on $r$.

With the aim of obtaining useful physics information from the MPD distribution, for each event we make a fit of the muon longitudinal development profile with the Gaisser-Hillas function~\cite{Gaisser:g-h}

\begin{equation}
\frac{\text{d}N}{\text{d}X} = \frac{\text{d}N_\mathrm{max}}{\text{d}X}\left(\frac{X-X_0}{\Xmum - X_0}\right)^{\frac{\Xmum-X_0}{\lambda}}e^{\frac{\Xmum-X}{\lambda}}. 
\label{eq:GH}
\end{equation}
Of the four parameters, \Xmum accounts for the point along the shower axis where the production of muons reaches its maximum as the shower develops through the atmosphere. As shown later, this parameter will be our main physics observable for composition studies. 

The best set of parameters that describes a given longitudinal muon profile (either at generation or reconstruction level) is obtained through a log-likelihood maximization of the Gaisser-Hillas function. When working with the MPD distributions at generation level (i.e., using the muon production points directly obtained from the simulation code \textsc{Corsika}~\cite{corsika}), we fit all of the Gaisser-Hillas parameters.

In reconstructed events, the number of muons used to build a MPD
distribution is not very large: after cuts typically {$\sim$}50 muons
({$\sim$}10 SD detectors) contribute for an energy of
10$^{19.5}$\,eV. Two reasons are at the source of this shortage: on
the one hand, the detectors are separated by 1.5\,km and have a finite
collecting surface of 8.7\,m$^{2}$ (the cross-section at
60$^{\circ}$); on the other hand, to minimize the distortions in the
reconstructed muon depths, we select tanks far from the core (and
therefore with small signals) to ensure an accurate determination of
\Xmum (see Sec.~\ref{s:deteffects} for further details). This muon
undersampling does not yield reliable estimates when all four
parameters of the Gaisser-Hillas are fitted. The solution adopted to
overcome this problem is to fix $X_0$ during the fitting
procedure. According to simulations, the preferred $X_0$ value depends
on the nature of the primary particle. Trying to avoid a
composition-dependent bias, we fix the parameter $X_\mathrm{0}$ to the
average value between the protons and iron nuclei, $X_0=-45\,{\rm g}/{\rm cm}^2$.  Assigning a particular value to $X_\mathrm{0}$ does not present a large source of systematic uncertainty given the weak correlation between \Xmum and $X_0$. We have observed that a shift of $10\,{\rm g}/{\rm cm}^2$ in $X_0$ translates into a variation of $1.5\,{\rm g}/{\rm cm}^2$ in the value of \Xmumm. As the mean difference in the $X_0$ values for proton and iron primaries is about $30\,{\rm g}/{\rm cm}^2$, a maximum bias of ${\sim} 3\,{\rm g}/{\rm cm}^2$ is expected. 
The MPD distribution fit is performed in an interval of depths ranging from 0 to $1200\,{\rm g}/{\rm cm}^2$, and it contains the entire range of possible values of \Xmum (the proton shower simulated\footnote{For each primary and hadronic model, 2000 \textsc{Corsika} simulations were used in this analysis.}  with the greatest depth of maximum has an energy of 96\,EeV and \Xmum $\approx1000\,{\rm g}/{\rm cm}^2$). 

\begin{figure}
\centering
\includegraphics[width=\columnwidth]{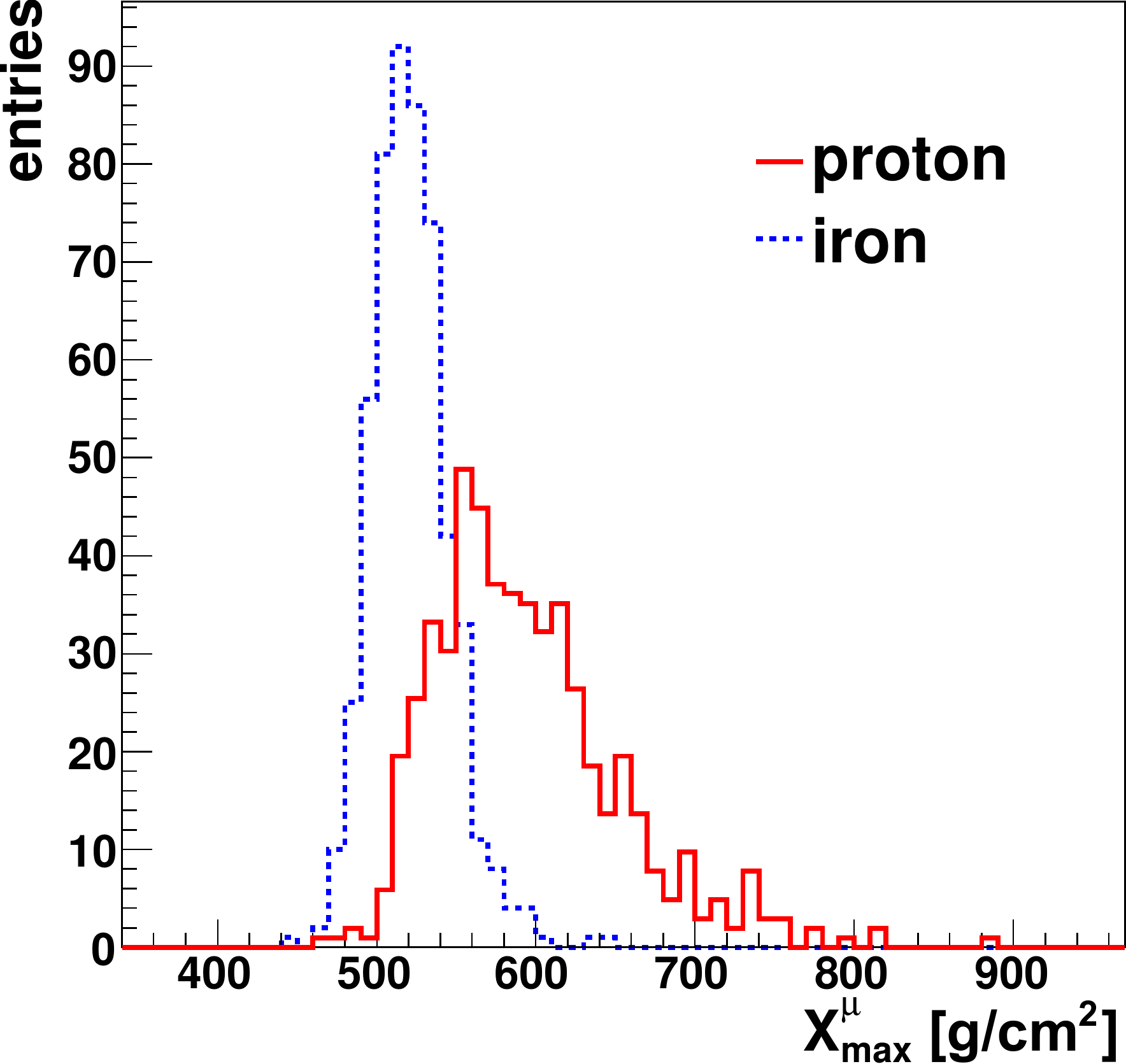}
\caption{\Xmum distributions for proton and iron showers simulated at
   30\,EeV with \Epos at zenith angles between 55$^{\circ}$ and 65$^{\circ}$. The mean value and the rms of the distributions show a clear dependence on the mass of the primary cosmic ray. For the construction of the MPDs, only muons reaching the ground at distances greater than 1700\,m were considered.
\label{fig:XmumaxDist}}
\end{figure}

From a sample of simulated proton and iron showers with 30\,EeV of energy, we observe  that the distribution of \Xmum varies as a function of the mass of the particle that initiates the atmospheric cascade (see Figure~\ref{fig:XmumaxDist}). For heavier particles, the average value of \Xmum is smaller and the distribution is narrower compared with that of lighter particles. The same behavior is observed when considering different energies for the primary particle. According to simulations, the \Xmum observable allows us to study the mass composition of UHECR data collected by a surface array of particle detectors.
In the following sections we investigate whether the systematic uncertainties associated with the Auger SD allow us to exploit the full physics potential associated with this observable.  
%
\section{Event Reconstruction}
\subsection{Detector effects}
\label{s:deteffects}
\begin{figure}
\centering
\includegraphics[width=0.49\columnwidth,height=1.5in]{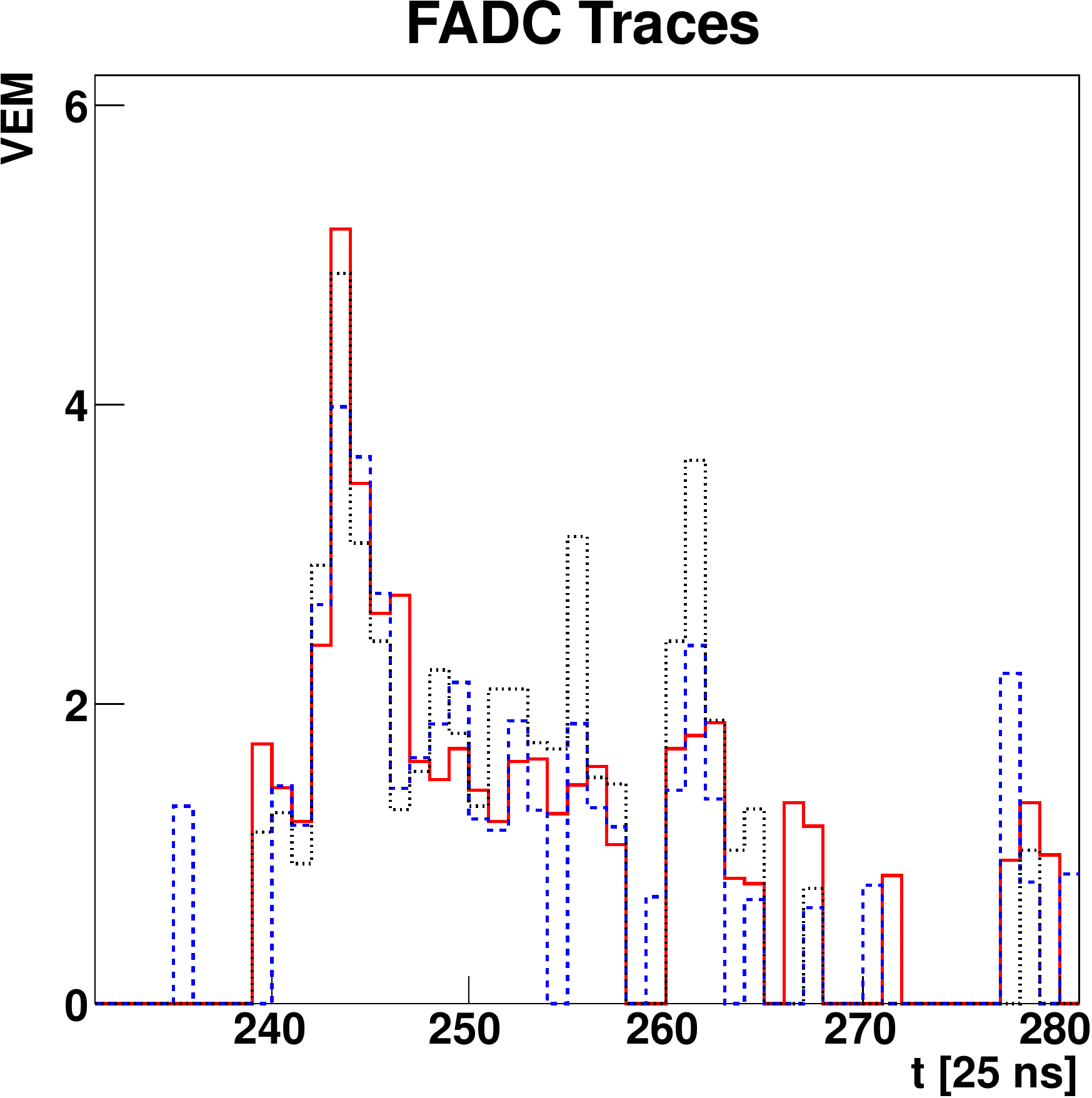}
\hfil
\includegraphics[width=0.49\columnwidth,height=1.5in]{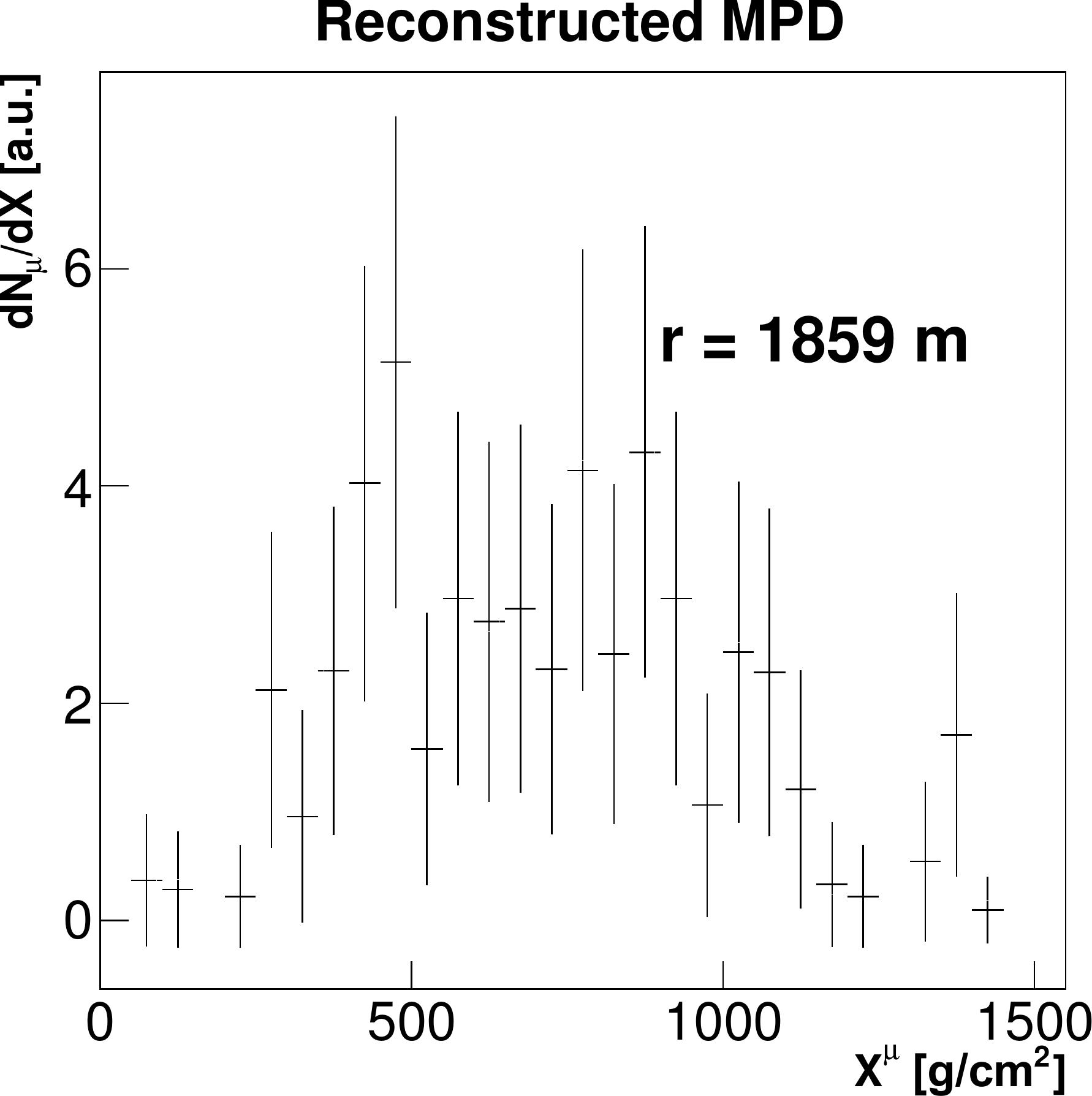}\\
\includegraphics[width=0.49\columnwidth,height=1.5in]{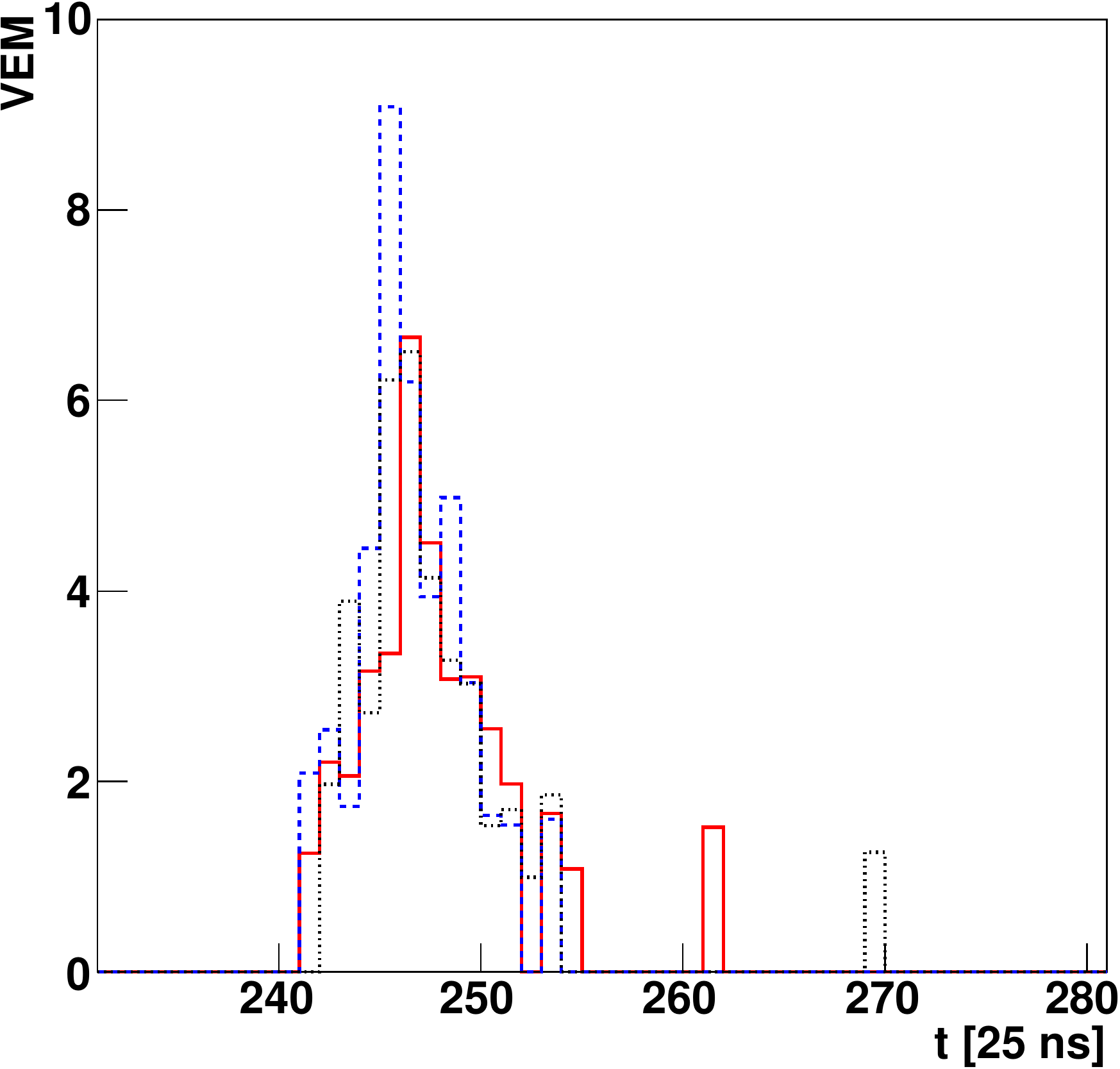}
\hfil
\includegraphics[width=0.49\columnwidth,height=1.5in]{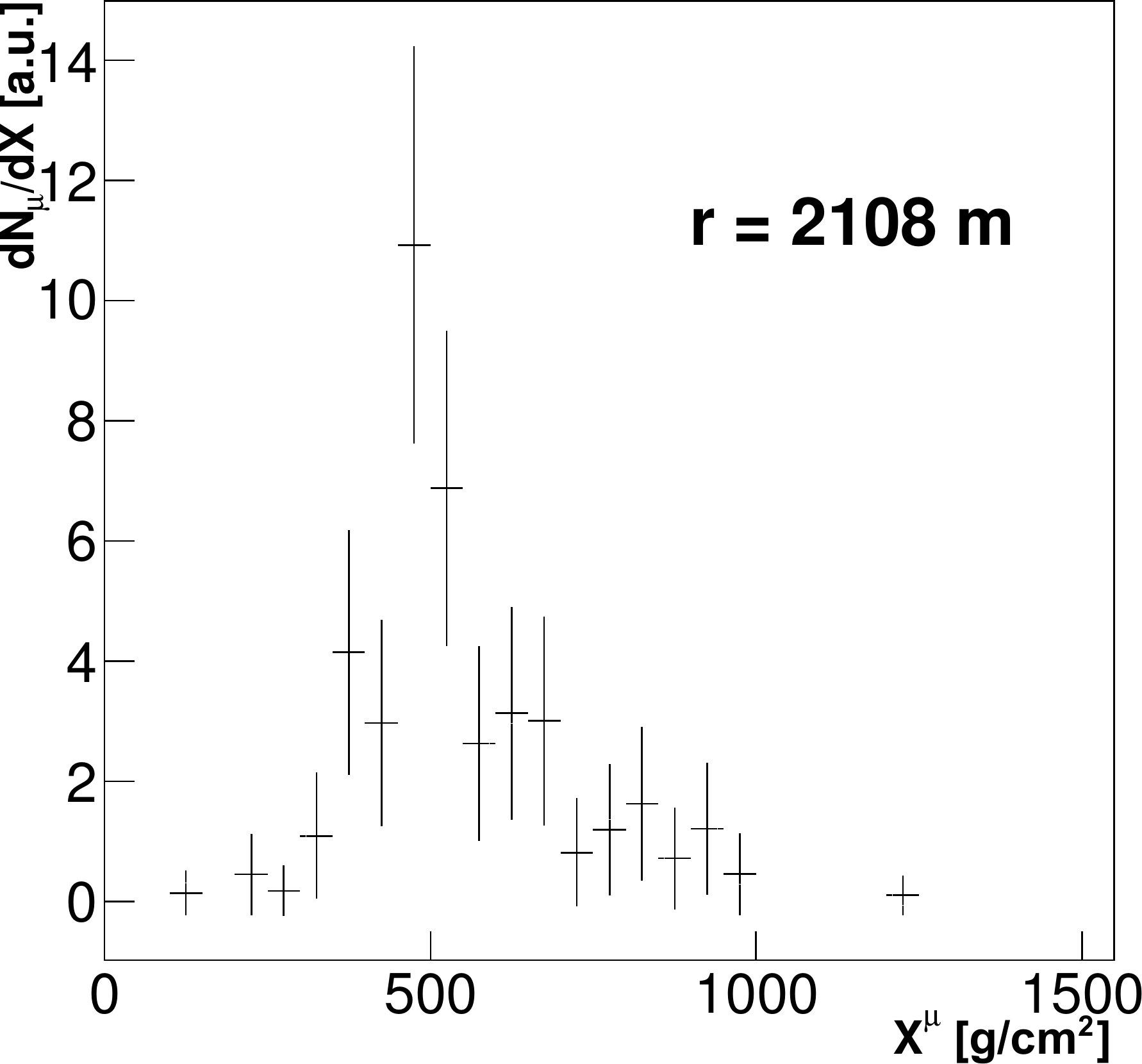}\\
\includegraphics[width=0.49\columnwidth,height=1.5in]{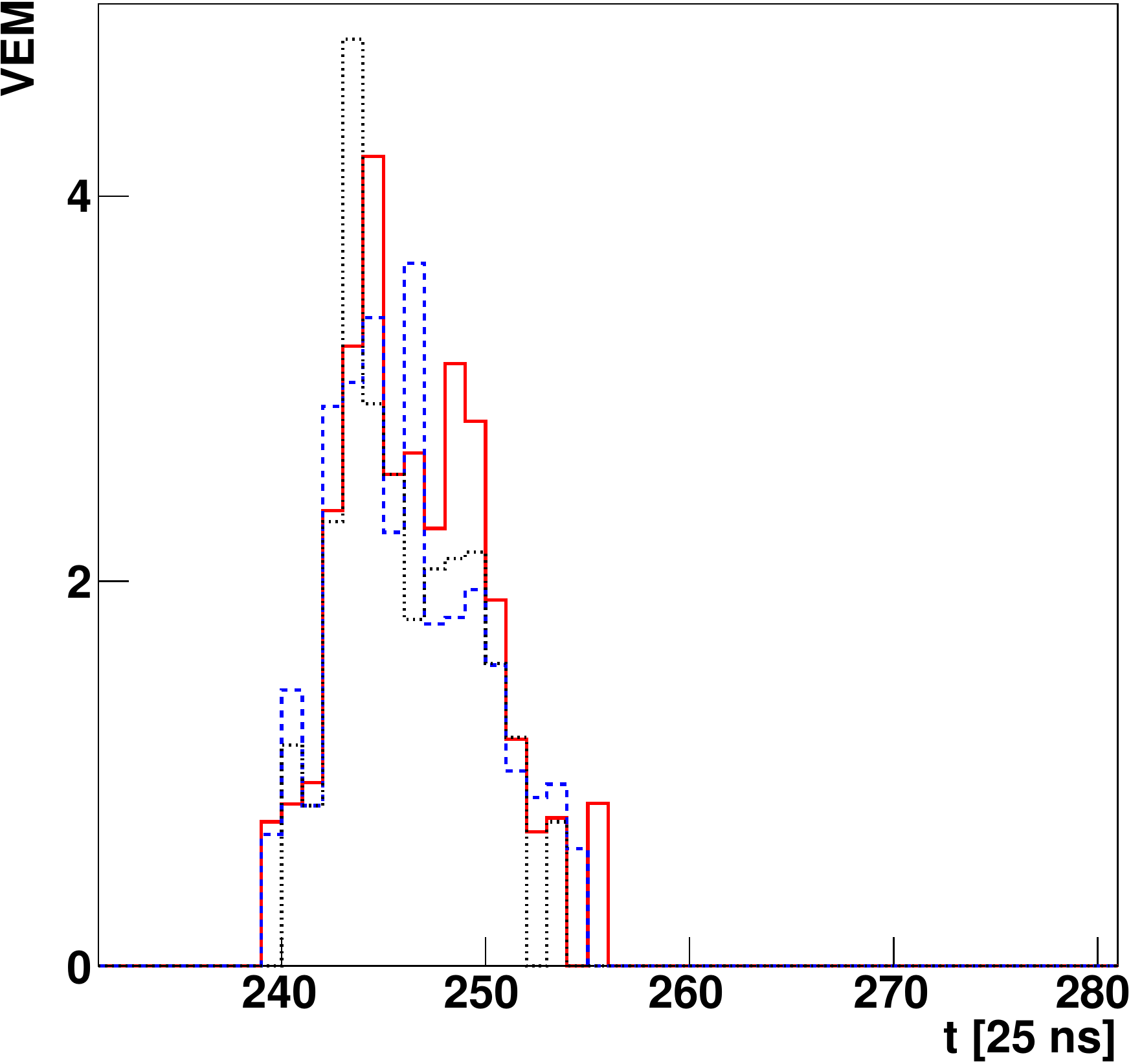}
\hfil
\includegraphics[width=0.49\columnwidth,height=1.5in]{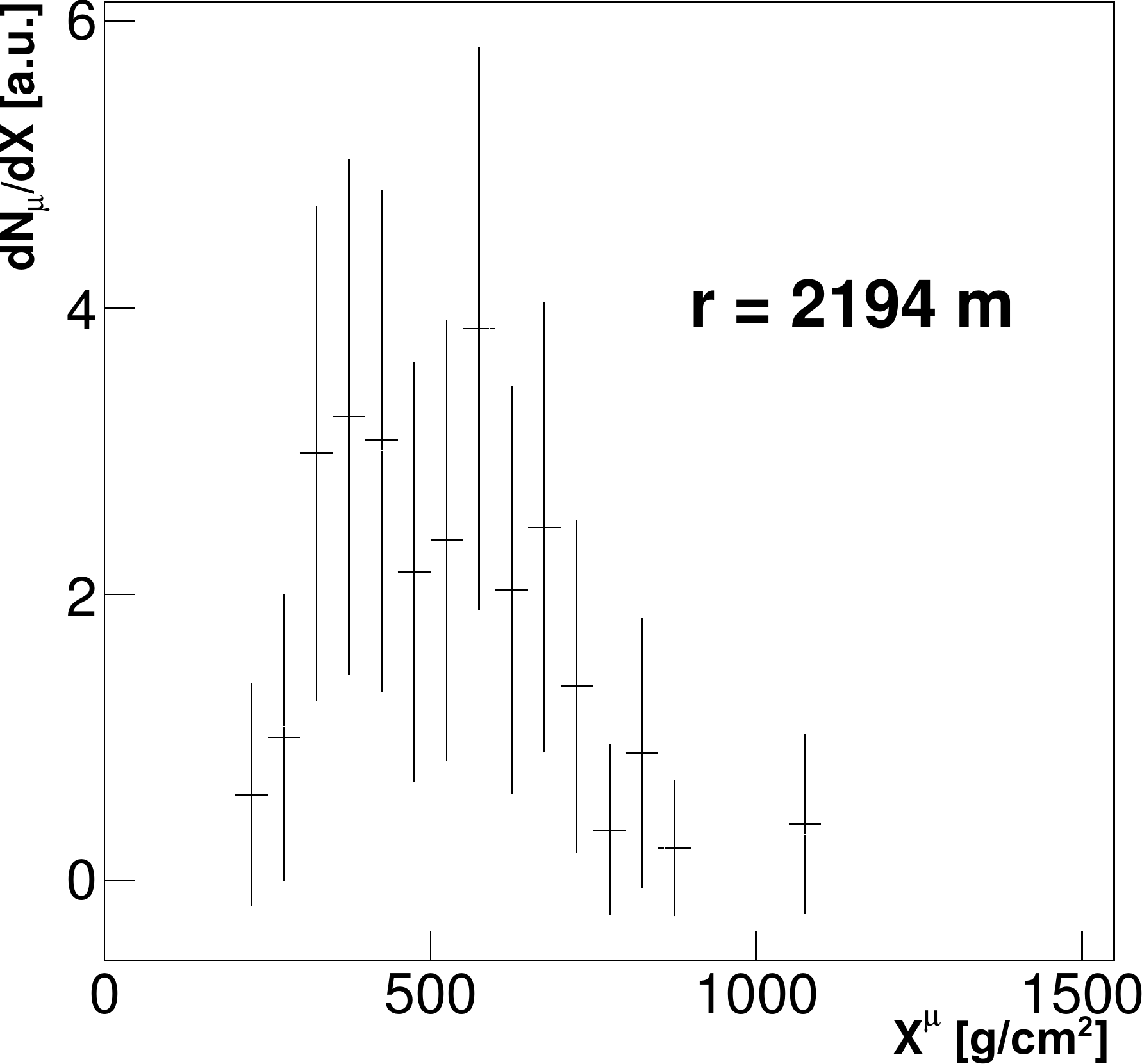}

\caption{\emph{Left:} Example of an event where FADC traces are shown
  at three different distances from the core. The signals recorded by
  the three PMTs appear in different colors. For the three stations,
  only the time bins contributing to the MPD reconstruction (signal
  $>$ 15\% of the maximum) are shown in the figure. \emph{Right:} Result of the
  conversion of the signals shown in the left panels into their
  corresponding MPDs. We show the distributions with a binning of
  $50\,{\rm g}/{\rm cm}^2$. For a thorough discussion of the full
  reconstruction see text. 
\label{fig:MPDrec}}
\end{figure}

The signals recorded by the Auger SD result from a mixture of muons
and electromagnetic (EM) particles. The reconstruction of the MPD
distribution for a given event requires the selection of the signal
solely due to muons. The EM signal is treated as a background that
must be eliminated. One way to achieve this is to work with inclined
showers (those having a zenith angle around or above 60$^{\circ}$). In
such showers, the EM component is heavily absorbed by the
atmosphere. The dependence of the MPD distribution shape with the
distance to the shower axis $r$ drastically decreases as $\theta$
increases, unlike the MPD reconstruction which worsens for increasing
values of $\theta$ [see Eq.~\eqref{eq:tres}]. Therefore, the
present work focuses only on data for which the zenith angles lie in
the interval [55$^{\circ}$, 65$^{\circ}$]. 

The EM contamination can be reduced even more by exploiting the different
behaviors of the EM and muonic components. In general, the EM signals are broader in time
and with smaller amplitudes. A cut on signal
threshold that rejects all time bins with signals below a certain
value ($S_\text{threshold}$) helps to diminish the EM
contamination.\footnote{Note that there is an additional EM component
that always accompanies muons and does not show such a strong
dependence with the distance to the core. This is sometimes referred
to as the EM halo, and comes from the decay of muons in flight. This
component is harder to avoid, but it follows more closely the time
distribution of the parent muons.} We set $S_\text{threshold}$ to 15\,\% of the maximum (peak)
of the recorded signal. This cut, apart from minimizing potential baseline
fluctuations, guarantees muon fractions above 85\,\%, regardless of the
energy and mass of the primary particle. 

For each entry in the MPD distribution, the uncertainty introduced in \Xmu ($\delta$\Xmus)
is a function of the time resolution ($\delta t$) and the accuracy of
reconstruction of the shower angle and core location. The uncertainty
in time gives rise to an uncertainty in the reconstruction of
$X^{\mu}$ that decreases quadratically with $r$ and increases with
$X^{\mu}$ as

\begin{equation}
\delta X^{\mu} = \frac{2X^{\mu} h_0}{r^2 \cos\theta}\text{ln}^{2}\left(\frac{X^{\mu} \cos\theta}{h_{0}\rho_0}\right) c\delta t.
\label{eq:tres}
\end{equation}
To derive Eq.~\eqref{eq:tres}, we have assumed an exponential atmospheric density $\rho(z) = \rho_{0}\exp(-z\cos\theta/h_0)$.
It is evident that the closer we get to the impact point
on the ground, the larger the uncertainty in \Xmus. Near the shower axis,
muons arrive closer in time, hence the impact of time resolution on
the estimation of the production depth is larger here than far from
the core. The contribution
of the geometric reconstruction to $\delta$\Xmu also increases
as we get closer to the core (with a linear dependence on $r$ in this case). Thus, to keep the distortions
of the reconstructed MPD small, only detectors far from the core are
useful. A cut in core distance, $r_\text{cut}$, is therefore mandatory.  This
cut diminishes the efficiency of the reconstruction, and it also affects the resolution
as it reduces the number of accepted muons: note that the
total uncertainty in the determination of the MPD maximum $\delta$\Xmum depends on the
number of muons $N_{\mu}$. However, the reconstruction efficiency improves with
energy, since the number of muons becomes larger as energy increases. Since
the number of muons at the ground level is a function of the mass of the
primary particle, in our selection we risk introduction of a bias towards
heavier nuclei, if the value for $r_\text{cut}$ is not carefully
chosen. Therefore, the selection of $r_\text{cut}$ must be a tradeoff
between the resolution of the reconstructed MPD distribution and the
selection bias~\cite{DGGThesis}. 
In particular, we require minimizing the uncertainty in the estimation of
\Xmu such that  (i) the reconstruction bias in \Xmum is smaller than
$10\,{\rm g}/{\rm cm}^2$, (ii) it is independent of the energy, and
(iii) there is a negligible (${<}\,10\%$) selection bias between primaries.

Using Monte Carlo simulations, we have found the optimal value for
$r_\text{cut}$ =  1700\,m which fulfills the above requirements, regardless of the
shower energy. Choosing an $r_\text{cut}$ which is independent of energy
implies that any difference in resolution for different
energies will be mainly a consequence of the different number of muons
detected at the ground. To estimate the impact of the sampling on the
determination of \Xmumm, we have studied how \Xmum changes as a
function of $r_\text{max}$ (the upper limit of
the distance interval [$r_\text{cut}$, $r_\text{max}$] used to integrate the
MPD profile). Our simulations show that variation of the \Xmum value
amounts to about $10\,{\rm g}/{\rm cm}^2$ per km shift in
$r_\text{max}$ (this effect is 3 times larger for events with
$\theta=45^{\circ}$). 
The fact that in the selected data we do not have triggered stations farther than
$\sim$4000\,m from the core implies that we build MPD distributions,
for both true and reconstructed levels,  by counting muons at the ground in
the distance range 1700\,m $< r <$ 4000\,m.  

To build the MPD distribution, every time bin of the 
FADC traces is converted into a MPD entry by means of
Eqs.~\eqref{eq:zcorr_t} and \eqref{eq:ztoX}. 
Muon arrival times are smeared by light propagation inside the
detector and the electronics response.
To compensate for this \textit{detector effect}, we subtract an offset
$t_\text{shift}$ to each time bin. Since in SD the raise time of the muon signal is much shorter than the consequent decay, the $t_\text{shift}$ depends on the $S_\text{threshold}$.  Simulating 
vertical muons that pass through an Auger surface detector, we find
that $t_\text{shift}$ = 73\,ns for $S_\text{threshold} = 15\%$, so this is the
correction we apply in our reconstruction. 

\begin{figure}
\centering

\includegraphics[width=\columnwidth,height=3in]{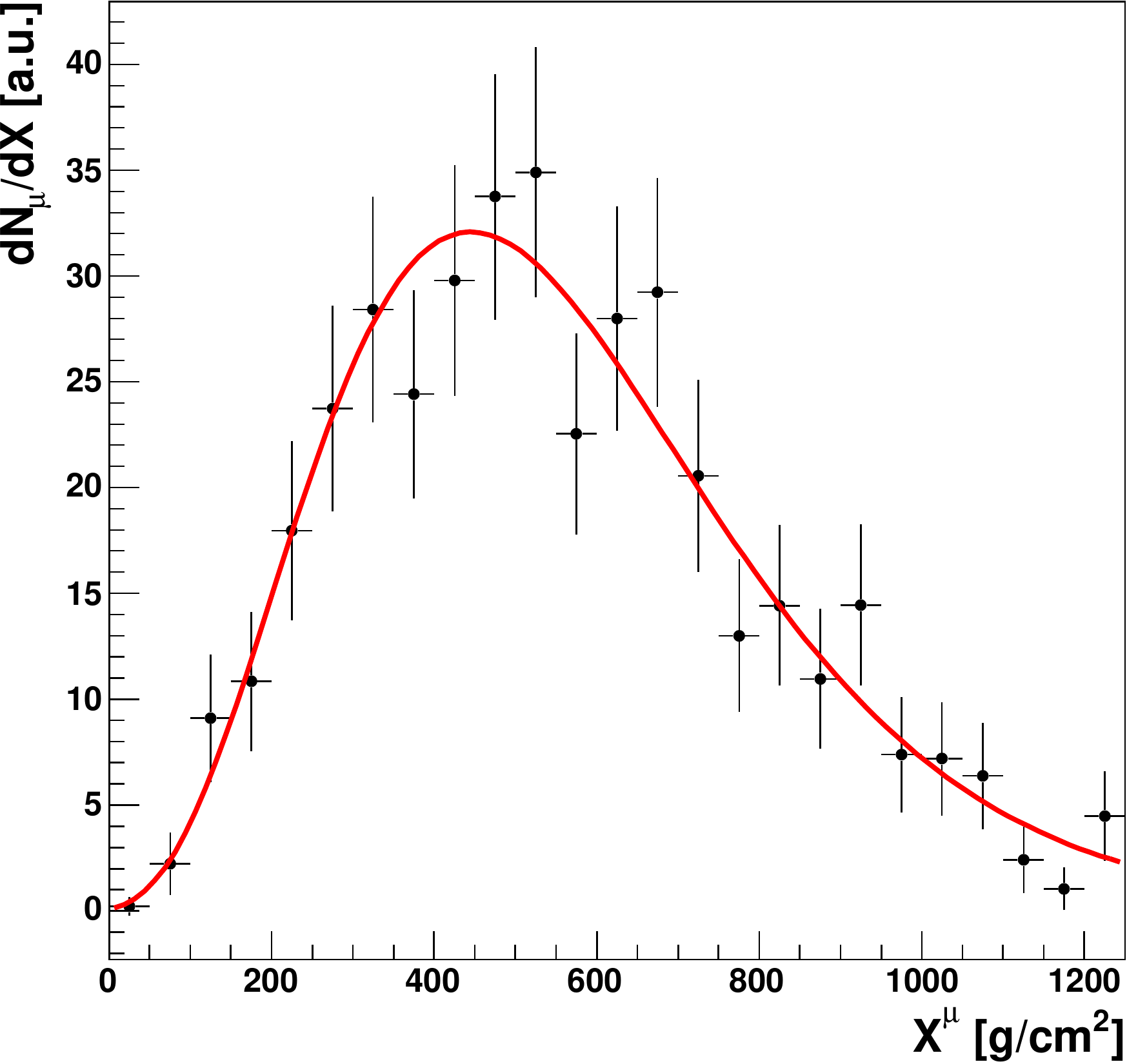}
\hfil
\includegraphics[width=\columnwidth,height=3in]{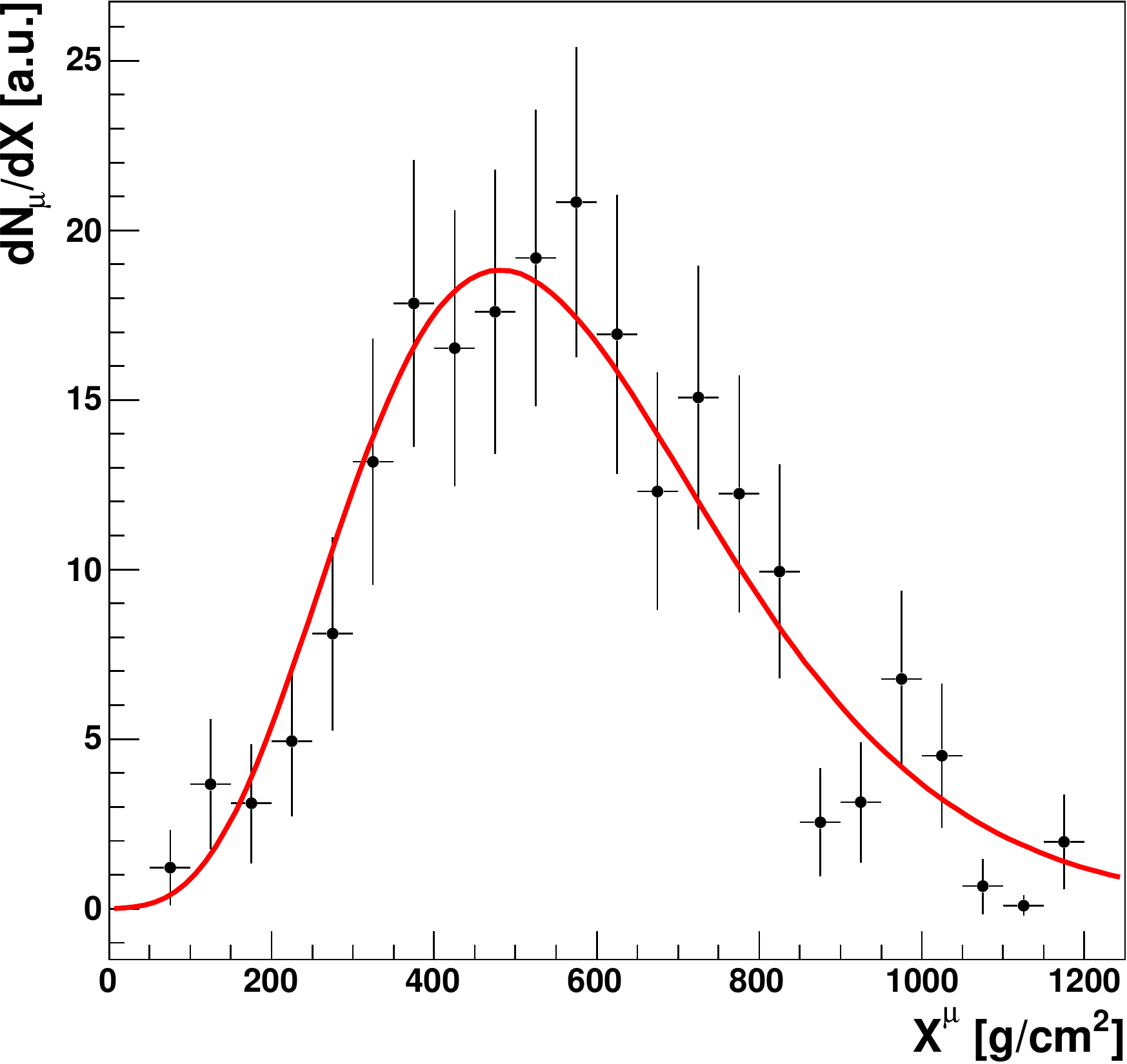}
\hfil
\includegraphics[width=\columnwidth,height=3in]{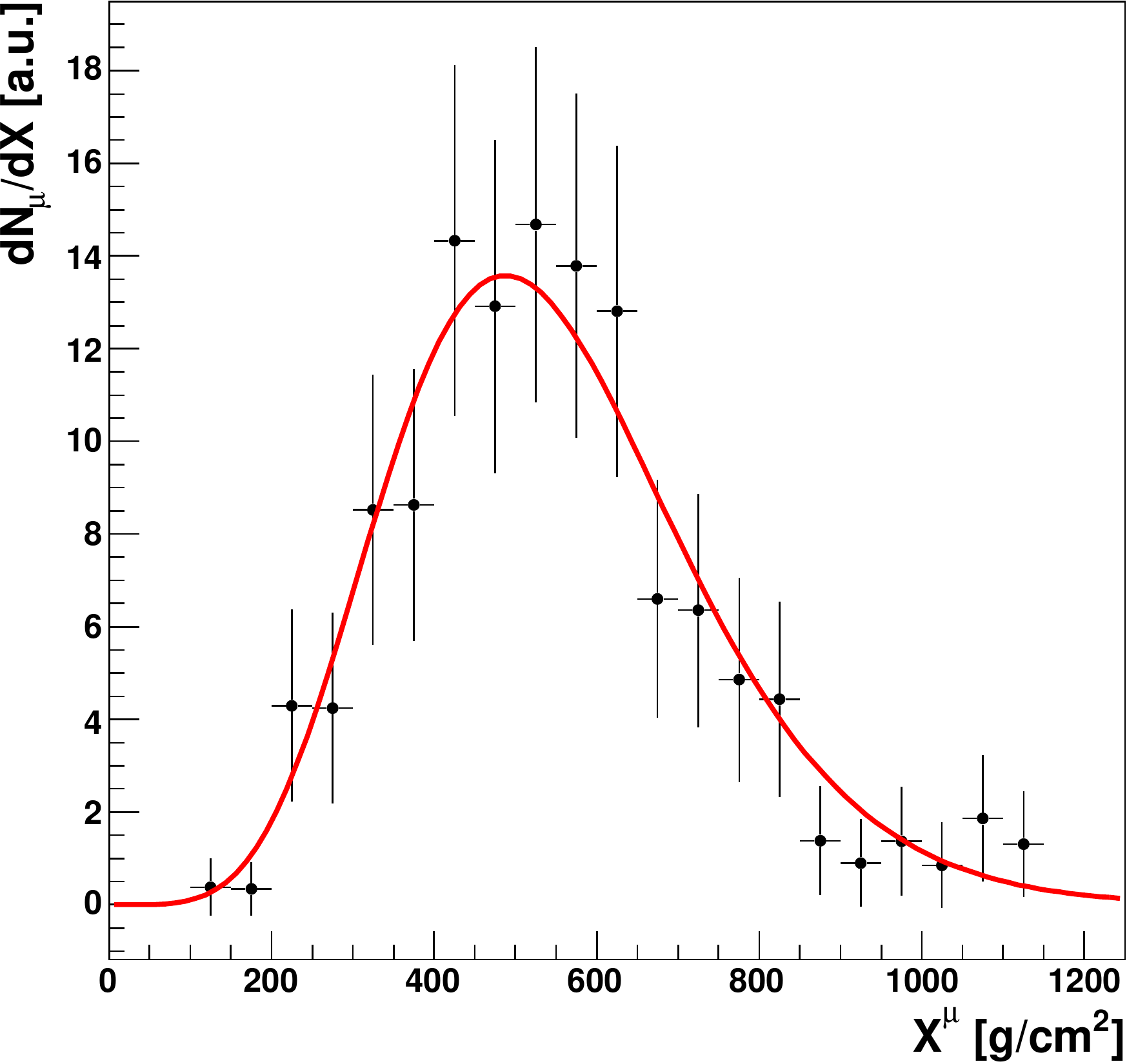}
\caption{Real reconstructed MPD distributions at three different
  energies: $E$ = (91 $\pm$ 3) EeV (top), $E$ = (33 $\pm$ 1) EeV
  (middle) and $E$ = (20 $\pm$ 1) EeV (bottom). The fits to a
   Gaisser-Hillas function are also shown.
\label{fig:realMPD}}
\end{figure}

The MPD distribution for a single detector is
obtained as the average of the MPD distributions yielded by each
working PMT.  Figure~\ref{fig:MPDrec} exemplifies how FADC traces are
mapped into individual MPDs. For each event, the final MPD profile is obtained by adding the individual
MPD distributions observed by each of the selected SD
stations. Figure~\ref{fig:realMPD} shows the reconstructed MPD
distribution for three real events at different energies.

\subsection{Selection criteria and resolution}
\label{s:cuts}
From the set of SD events with a reconstructed MPD distribution, we select 
those with reliable measurements of longitudinal profiles. This
requires the application of a simple set of selection criteria which
are described below:
\begin{description}
\item[Energy cut.] Since the number of muons increases with the energy
  of the primary, 
we have observed that in
events with energies below 20 EeV the number of entries in the
resulting MPD is very small, giving a very poor determination of the
\Xmum observable. In fact, below this energy cut the number of
selected muons can drop below 10. Therefore, we restrict
our analysis to events with energy larger than 20 EeV.
\item[$\mathbf{X^{\boldsymbol\mu}_\mathrm{\mathbf{max}}}$ uncertainty.]
  We reject events whose relative uncertainty $\delta$\Xmumm/\Xmum is larger than a certain value
  $\epsilon_\mathrm{max}$. This upper limit is an energy-dependent
  quantity (see Table~\ref{t:epsilon}) since  the accuracy in the
  estimation of \Xmum improves with energy. This is a natural
  consequence of the increase in the number of muons that enter the
  MPD distribution as the energy grows.  
\end{description}
\begin{table}[!t]
\caption{Maximum relative uncertainties allowed in the estimation of
  \Xmumm. The value chosen for $\epsilon_\mathrm{max}$ ensures no
  selection bias between the different primary species.}  
\centering
\begin{tabular}{c c}
\toprule
log$_{10}$(E/eV) &  $\epsilon_\mathrm{max}$[\%] \\ 
\colrule
$[$19.3, 19.4$]$ & 15 \\
$[$19.4, 19.6$]$ & 11 \\
$[$19.6, 19.7$]$ & 10 \\
$[$19.7, 19.8$]$ & 8 \\
$>$ 19.8 & 7 \\
\botrule
\end{tabular}
\label{t:epsilon}
\end{table}

\begin{figure}[!t]
 \vspace{5mm}
 \centering
\includegraphics[width=\columnwidth]{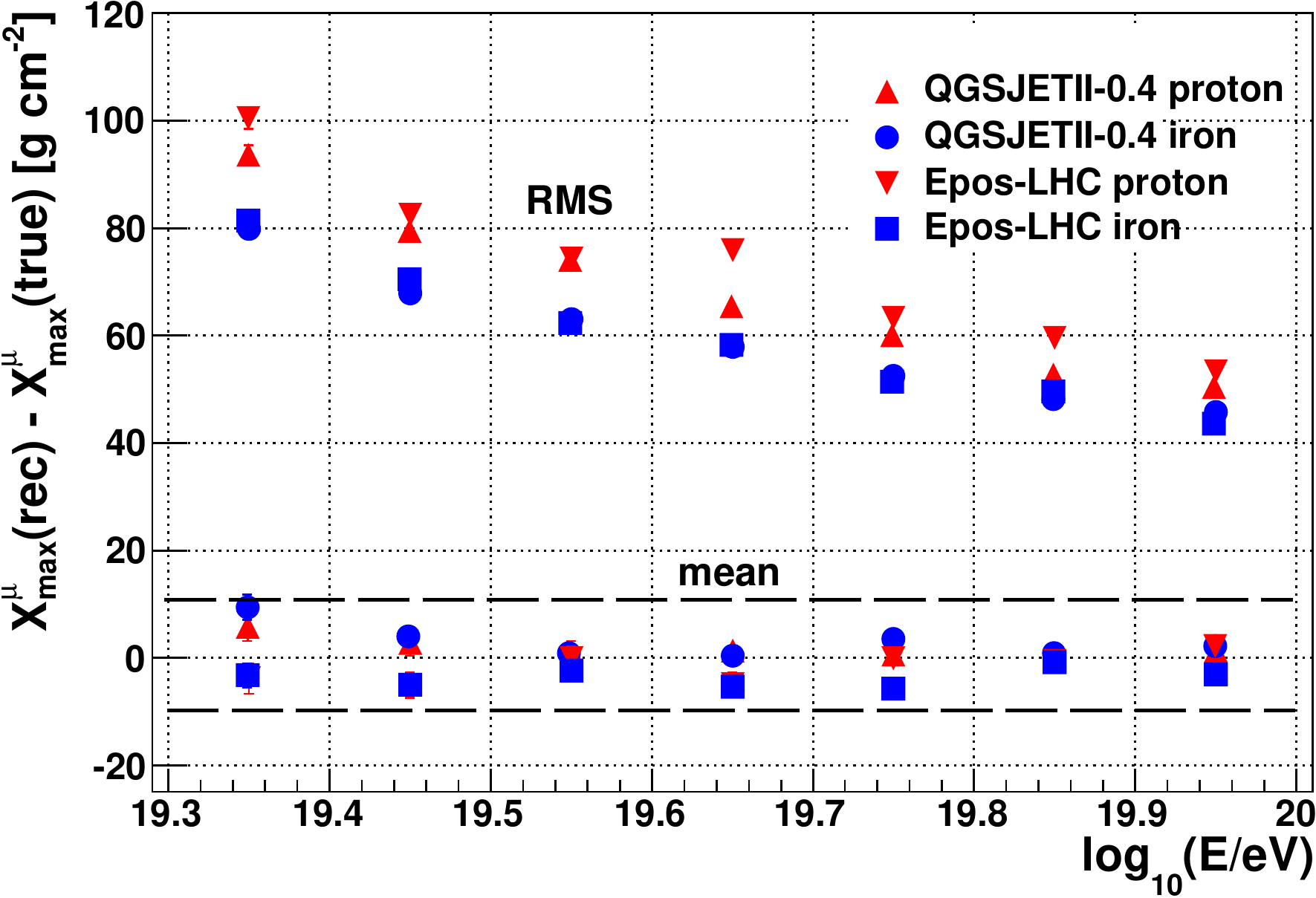}
 \caption{Evolution with energy of the mean and rms of the distribution
   \Xmumm(reconstructed) $-$ \Xmumm(true). The simulations were made using
the \QII~\cite{QII} and \Epos hadronic models for protons
and iron nuclei for $55^\circ\leq \theta\leq 65^\circ$. Dashed lines
indicate the final systematic uncertainty bounds due to
the reconstruction effects, different hadronic models, and primary particles.}
 \label{fig:resol}
\end{figure}

For simulated events, the overall selection efficiency grows from 85\%
(at 20\,EeV) to almost 100\% (for energies larger than 40\,EeV). Monte
Carlo studies show that the chosen cuts introduce a negligible composition
bias (smaller than $2\,{\rm g}/{\rm cm}^2$). 
As shown in Figure~\ref{fig:resol}, the absolute value of the
mean bias after reconstruction is ${<}\,10\,{\rm g}/{\rm cm}^2$, 
regardless of the hadronic model, energy, and atomic mass of the
simulated primary particle.
The resolution, understood as the rms of the distribution
\Xmumm(reconstructed) $-$ \Xmumm(true),
ranges from 100 (80)$\,{\rm g}/{\rm cm}^2$ for a proton (iron) at the
lower energies to about $50\,{\rm g}/{\rm cm}^2$ at the highest energies
(see Figure~\ref{fig:resol}). The improvement of the resolution with
energy is a direct consequence of the increase in the number of 
muons.

Several sources contribute to the total resolution in the measurement
of \Xmumm. Based on simulations, we have estimated the importance of
each of them in this particular analysis. The largest
contribution comes from the number of selected muons, which accounts
for more than 50$\%$ of the total resolution. Of negligible importance
(below 1$\%$) is the contribution due to the method itself (namely,
the kinematic delay approximation, with an average fixed energy per
muon, instead of the true unknown energy of each muon). 
The influence of the time uncertainty and the accuracy
of the geometric reconstruction of the shower are at the levels of
30$\%$ and  15$\%$, respectively. We have been able to minimize their
impact on the \Xmum measurement with the dedicated analysis,
optimizing the values of $r_\text{cut}$ and $S_\text{threshold}$. 

An important point to be discussed is how our analysis is affected by the observation
that the number of muons in simulations is smaller than the one
measured in data~\cite{AugerMuons, AugerMuons2}. If the discrepancy is naively
reduced just to a normalization problem, then only the MPD resolution
is altered. In such a case, the expected resolution improves when the
number of muons is augmented in simulated events. As discussed
above, the number of muons is the largest contribution to the measured
resolution. However, if the muon number is not only scaled up, but
the muon energy spectrum is also modified~\cite{Ulrich:2010rg}, then
the kinematic delay needs reevaluation. Since we select stations located far away from the core
location ($r_\text{cut} =$ 1700\,m), the kinematic delay is not a dominant
effect (see Figure~\ref{fig:Delays}). Therefore the conclusions of
this work are unaltered.
\subsection{Systematic uncertainties}
A careful examination of the systematic effects influencing our measurements has been carried out. Below, we list and quantify the most relevant sources contributing to the overall systematic uncertainty:
\begin{description}
\item[Reconstruction, hadronic model and primary mass.] As can be seen in Figure~\ref{fig:resol}, the difference between the generated and the reconstructed muonic shower maximum is bracketed by  $\pm\,10\,{\rm g}/{\rm cm}^2$ for proton and iron primaries with energies above 20\,EeV. We take this value as an estimate of the overall systematic uncertainty due to the reconstruction effects, differences in the hadronic interaction models, and differences due to the unknown nature of the primary particle.
\item[Seasonal effect.] The data show a dependence of the measured \Xmum value with seasons; e.g., in summer we measure deeper \Xmum values than in winter. Fitting a sinusoidal function to this dependency, the amplitude is $12\pm2\,{\rm g}/{\rm cm}^2$. Monte Carlo events generated for different seasons do not show such a modulation (either at generation or after detector effects are considered). Several tests probed unsuccessfully to identify the source of this discrepancy, and therefore we include the amplitude of this effect as a systematic uncertainty.
\item[Time variance model.] The uncertainty on the arrival time of the
  EAS front has been modeled from the data. It influences the
  reconstruction of the curvature and of the impact point on the
  ground, and it has a direct impact on the reconstruction of the
  maximum of the muon production depth. To obtain the associated
  systematic uncertainty introduced by this model we compared two
  different parameterizations of the time variance~\cite{ALS:2007}
  model. They both have a common contribution from the resolution on
  the absolute time given by the GPS and from the 40\,MHz sampling of
  the FADCs, and they differ in the modeling of the fluctuations of the arrival time of the first particle. The difference between the two models induces a $5\,{\rm g}/{\rm cm}^2$ systematic uncertainty on the determination of the maximum of the muon production depth.
\item[Accidental signals.]
In real events, a background of random accidental signals might appear. The most frequent source of random noise is created by single particles (generally isolated atmospheric muons) and, more rarely, by a bunch of particles arriving at the same time from a low-energy shower close to a SD station. In general, it is very difficult to identify and take into account all possible sources of accidental signals. They can appear at any time and at any location in the SD array, completely uncorrelated with the genuine primary shower signal. Random accidental signals can have a damaging effect on the data quality, since they can trigger some stations of the array, distorting the reconstruction of the showers. In our analysis, the main impact comes from a possible underestimation of the start time of the traces due to an accidental signal prior to the true one. Using an unbiased sample of random accidental signals extracted from data events collected in the SD stations, we have studied the influence of accidental signals in the Monte Carlo reconstructions. Regardless of the energy and primary mass, we have found a systematic underestimation by ${\sim}4.5\,{\rm g}/{\rm cm}^2$ in the determination of \Xmumm. We have corrected for this bias in our data.
\item[Atmospheric profile.] For the reconstruction of the MPD profiles, the atmospheric conditions at the Auger site, mainly height-dependent atmospheric profiles, have to be well known. To quantify the influence of the uncertainty in the reconstructed atmospheric profiles on the value of \Xmumm, a direct comparison of GDAS data\footnote{GDAS is a publicly available data set containing all main state variables dependent on altitude with a validity of 3 hours for each data set~\cite{GDAS, GDASpao}.} with local atmospheric measurements\footnote{Intermittent meteorological radio soundings with permanent ground-based weather stations.} has been performed on an event-by-event basis.  We have obtained a distribution with a small shift of $2.0\,{\rm g}/{\rm cm}^2$ in \Xmum and a rms of $8.6\,{\rm g}/{\rm cm}^2$.
\item[Selection efficiency.] The selection efficiency for heavy primaries is larger than for protons since the former have a muon-richer signal at the ground. The analysis was conceived to keep this difference below 10$\%$ for the whole energy range. This difference in efficiency, although small, may introduce a systematic effect in the determination of \Xmumm. We have determined it by running our analysis over a 50/50 mixture of protons and iron, resulting in a negligible contribution to the systematic uncertainty of $\le2\,{\rm g}/{\rm cm}^2$.
\end{description}

Table~\ref{t:sys} summarizes the sources contributing to the
systematic uncertainty. The overall systematic uncertainty 
in $\langle$\Xmumm$\rangle$ amounts to ${\sim}17\,{\rm g}/{\rm cm}^2$. This represents approximately 25\% of the proton-iron separation.

\begin{table}[!t]
\caption{Evaluation of the main sources of systematic uncertainties in \Xmumm.} 
\centering
\begin{tabular}{c c}
\toprule
Source & Sys.\,uncertainty [${\rm g}/{\rm cm}^2$] \\ 
\colrule
Reconstruction, & \multirow{ 2}{*}{10}\\ hadronic model and primary\\ 
Seasonal effect & 12 \\
Time variance model & 5 \\[1ex] 
\colrule
\bf{ Total} & \bf{17}\\
\botrule
\end{tabular}
\label{t:sys}
\end{table}
%
\section{Results}
\label{s:results}
The data set used in this analysis comprises events
recorded between 1 January 2004 and 31 December 2012. We compute the MPD distributions on
an event-by-event basis. 
To guarantee an accurate reconstruction of the longitudinal profile, we
impose the selection criteria described in Sec.~\ref{s:cuts}. For
the angular range and energy threshold set in this analysis, our initial
sample contains 500 events. After our quality cuts, it is reduced to 481
events. 

\begin{figure}[!t]
 \vspace{5mm}
 \centering
\includegraphics[width=\columnwidth]{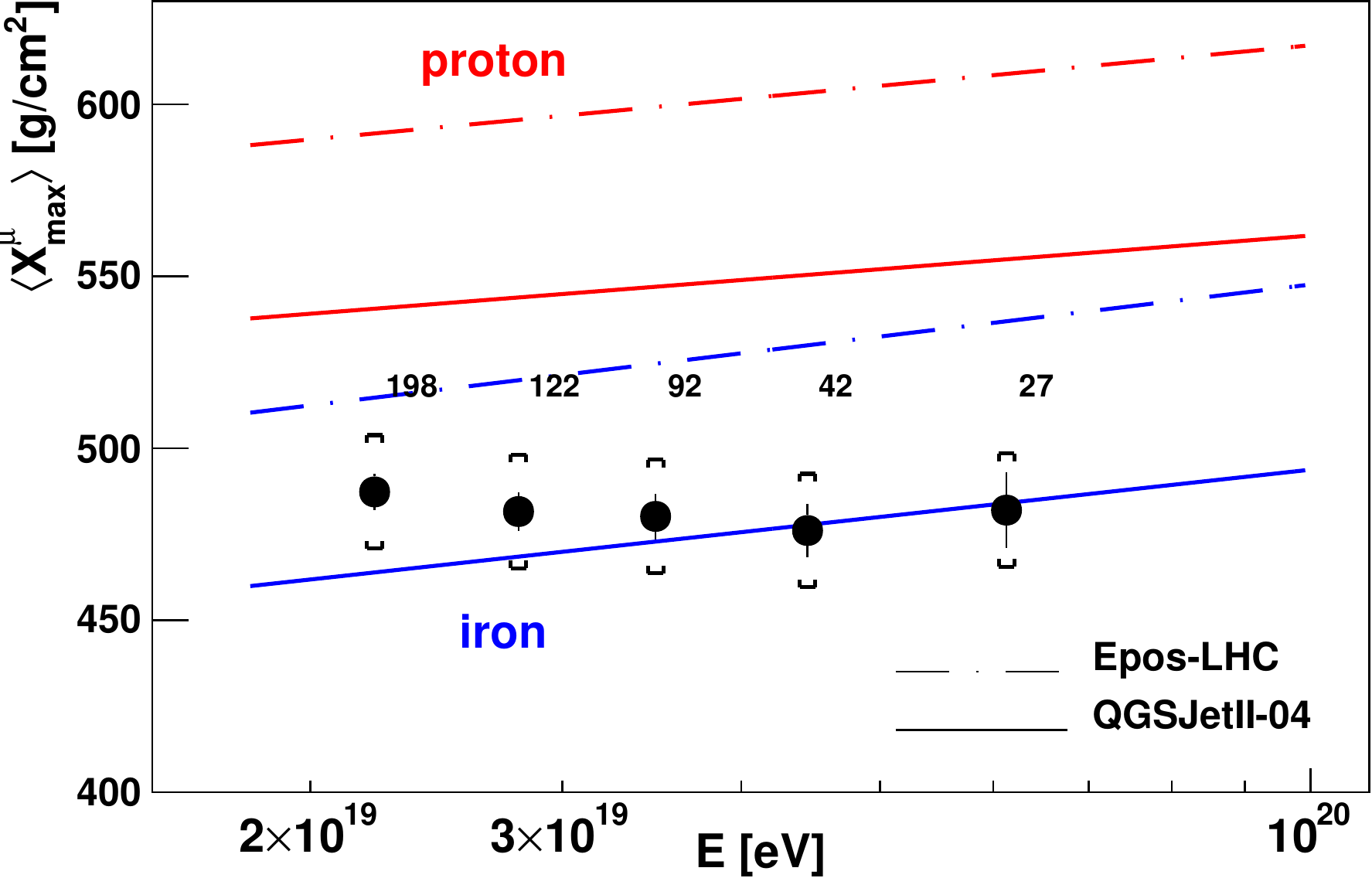}
 \caption{$\langle$\Xmumm$\rangle$ as a function of energy. The
   predictions of different hadronic models for protons and
   iron are shown. Numbers indicate the number of events in each
   energy bin, and brackets represent the systematic uncertainty.}
 \label{fig:ER}
\end{figure}

The evolution of the measured $\langle$\Xmumm$\rangle$ as a function
of the energy is shown in Figure~\ref{fig:ER}. The data are grouped
in five energy bins of width 0.1 in $\log_{10}(E/\text{eV})$, except for the
last bin, which contains all events with energy above $\log_{10}(E/\text{eV})=19.7$
($E$ = 50\,EeV). The sizes of error bars represent the standard deviation of the mean. 

%
\section{Discussion}
\begin{figure*}[!t]
 \centering
\includegraphics[width=\textwidth]{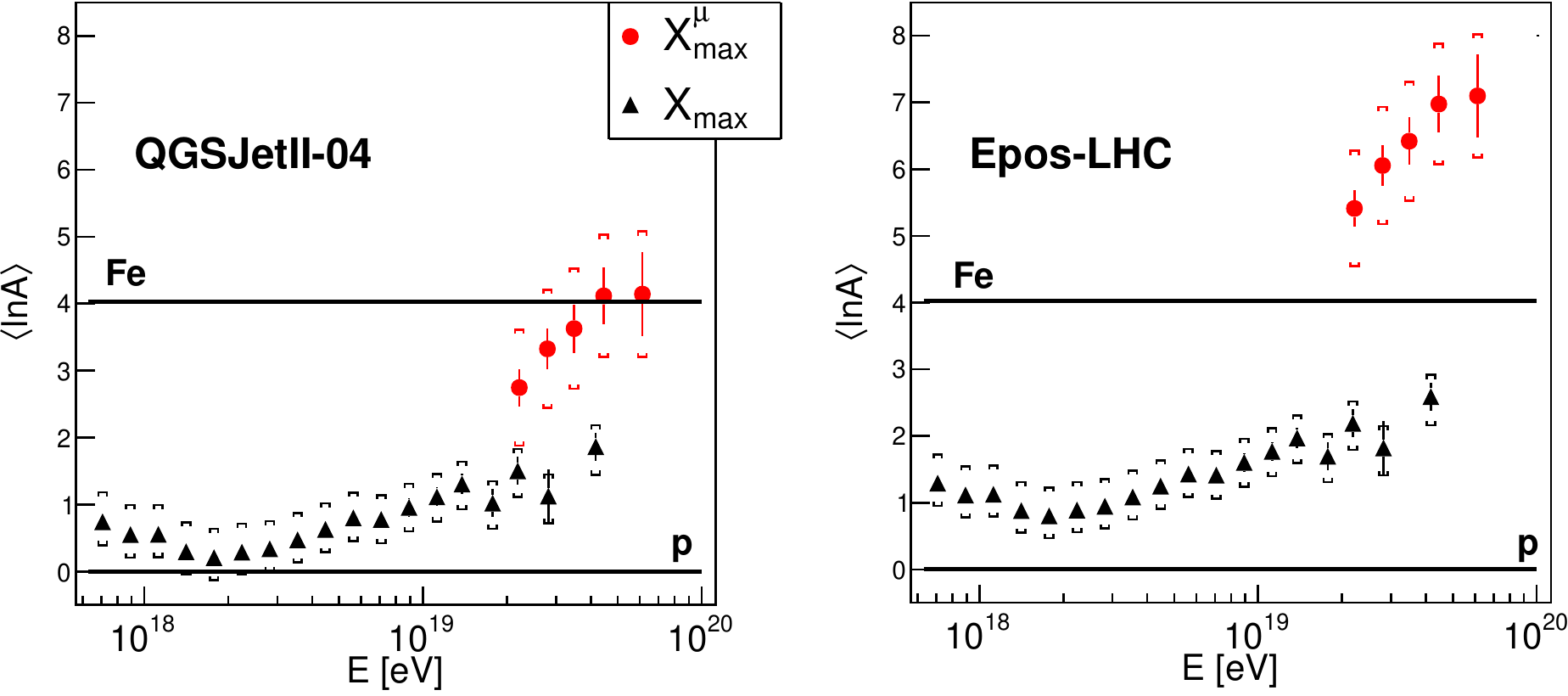}
 \caption{Conversion of $\langle\Xmumm\rangle$ (circles) and
$\langle\Xmm\rangle$ (triangles)~\cite{antoine} to $\langle\ln A\rangle$, as a
function of energy. On the left (right) plot we use \QII
(\Epos) as the reference hadronic model. See text for a detailed
discussion of the difference between models. Brackets correspond
to the systematic uncertainties.}
 \label{fig:lnA}
\end{figure*}

Under the assumption that air-shower simulations are a fair
representation of reality, we can compare them to data in order to
infer the mass composition of UHECRs. For interaction models (like
those used for Figure~\ref{fig:ER}) that assume that no new physics
effects appear in  hadronic interactions at the energy scales probed
by Auger, the evolution of the mean \Xmum values indicates a change in
composition as the energy increases. Data show a flatter trend than
pure proton or pure iron predictions ($35.9\pm1.2$ and
$48.0\pm1.2\,{\rm g}/{\rm cm}^2/{\rm decade}$,
respectively\footnote{Mean values between \QII and \Epos predictions.}). We measure a value of ${\rm d}\langle\Xmumm\rangle/{\rm
  d}\log_{10}E=-25\pm22\,(\text{stat.})\pm21\,(\text{syst.})\,{\rm
  g}/{\rm cm}^2/{\rm decade}$. This value deviates from a pure proton
(iron) composition by 1.8 (2.3)\,$\sigma$. 

In Figure~\ref{fig:ER} we observe how \QII and \Epos estimate, for
both protons and iron, a similar {\it muonic elongation rate}
(evolution of \Xmum with energy) but with considerable differences in
the absolute value of \Xmumm. While the Auger data are bracketed by
\QII, they fall below the \Epos estimation for iron. Therefore, the
study of the MPD profile can also be used as a tool to constrain
hadronic interaction models. 

\Xm and \Xmum are strongly correlated, mainly by the depth of first
interaction~\cite{DGGThesis, Ruben}. According to simulations the
correlation factor between these two observables is
$\ge$\,0.8. Therefore, similarly to \Xmm, \Xmum is correlated with the
mass of the incident cosmic ray particle. We can thus convert both
observables into $\langle\ln A\rangle$ using the same interaction
model~\cite{massJCAP,  Anh}. 

Figure~\ref{fig:lnA} shows the outcome of this conversion for two
different hadronic models. For \Epos the results indicate primaries
heavier than iron ($\ln A_\text{Fe}\simeq4$). The mean $\ln A$ values
extracted from the measurements of \Xm and \Xmum are incompatible with each other at a
level of at least 6\,$\sigma$. \Epos in combination with \Fluka
as a low-energy interaction model does not offer a consistent
description of the electromagnetic and muonic components of the
EAS. With \QII/\textsc{Fluka}, we obtain compatible values for $\ln A$
within 1.5\,$\sigma$, but it should be noted that, in contrast to
\Epos, this model has problems to describe in a consistent way the
first two moments of the $\ln A$ distribution obtained from the \Xm
measurements done with the FD~\cite{massJCAP}.  We conclude from the
comparisons shown in Ref.~\cite{massJCAP} and here that none of the
interaction models recently tuned to LHC data provide a consistent
description of the Auger data on EM and MPD profiles. 

The found discrepancies underline the complementarity of the
information provided by the longitudinal profiles of the
electromagnetic particles and the muons.  The EM profile in a shower
originates mainly from the decay products of high-energy neutral pions
produced in the first few interactions and is thus closely related to
the features of hadronic interactions at \emph{very high} energies. In
contrast, the MPD profile is an integral measure of \emph{high} and
\emph{intermediate} energy interactions, as most charged pions decay
only once they have reached energies below 30\,GeV. While details
of interactions at a few 100\,GeV are insignificant for the EM profile,
they are of direct relevance to muons. Hence, the measurement of muon
profiles provides valuable insight that sets additional constraints on
model descriptions and will help to improve our understanding of
hadronic interactions.  
\section{Conclusions}
The FADC traces from the water Cherenkov detectors of the Pierre Auger
Observatory located far from shower cores have been used to make
a reconstruction of the muon production depth distribution on an
event-by-event basis. The maximum of the distribution \Xmum contains
information about the nature of UHECRs. However, the current
level of systematic uncertainties associated with its determination 
prevents us from making conclusive
statements on mass composition. We have also discussed 
how \Xmum allows for a direct test of
hadronic interaction models at the highest energies, thus showing the
power of UHECR data to probe fundamental interactions in an energy
regime well beyond those reached at LHC. 
This analysis has established a novel approach to study the
longitudinal development of the hadronic component of EASs. 
\section*{Acknowledgments}
The successful installation, commissioning, and operation of the Pierre Auger Observatory would not have been possible without the strong commitment and effort from the technical and administrative staff in Malarg\"{u}e. 

We are very grateful to the following agencies and organizations for financial support: 
Comisi\'{o}n Nacional de Energ\'{\i}a At\'{o}mica, Fundaci\'{o}n
Antorchas, Gobierno De La Provincia de Mendoza, Municipalidad de
Malarg\"{u}e, NDM Holdings and Valle Las Le\~{n}as, in gratitude for
their continuing cooperation over land access, Argentina; the
Australian Research Council; Conselho Nacional de Desenvolvimento
Cient\'{\i}fico e Tecnol\'{o}gico (CNPq), Financiadora de Estudos e
Projetos (FINEP), Funda\c{c}\~{a}o de Amparo \`{a} Pesquisa do Estado
de Rio de Janeiro (FAPERJ), S\~{a}o Paulo Research Foundation (FAPESP)
Grants \# 2010/07359-6, \# 1999/05404-3, Minist\'{e}rio de Ci\^{e}ncia
e Tecnologia (MCT), Brazil; MSMT-CR LG13007, 7AMB14AR005,
CZ.1.05/2.1.00/03.0058 and the Czech Science Foundation grant
14-17501S, Czech Republic;  Centre de Calcul IN2P3/CNRS, Centre
National de la Recherche Scientifique (CNRS), Conseil R\'{e}gional
Ile-de-France, D\'{e}partement Physique Nucl\'{e}aire et Corpusculaire
(PNC-IN2P3/CNRS), D\'{e}partement Sciences de l'Univers
(SDU-INSU/CNRS), Institut Lagrange de Paris, ILP LABEX ANR-10-LABX-63,
within the Investissements d'Avenir Programme  ANR-11-IDEX-0004-02,
French National Research Agency under grant no.\ ANR-2010-COSI-002, France; Bundesministerium f\"{u}r Bildung und Forschung (BMBF), Deutsche Forschungsgemeinschaft (DFG), Finanzministerium Baden-W\"{u}rttemberg, Helmholtz-Gemeinschaft Deutscher Forschungszentren (HGF), Ministerium f\"{u}r Wissenschaft und Forschung, Nordrhein Westfalen, Ministerium f\"{u}r Wissenschaft, Forschung und Kunst, Baden-W\"{u}rttemberg, Germany; Istituto Nazionale di Fisica Nucleare (INFN), Ministero dell'Istruzione, dell'Universit\`{a} e della Ricerca (MIUR), Gran Sasso Center for Astroparticle Physics (CFA), CETEMPS Center of Excellence, Italy; Consejo Nacional de Ciencia y Tecnolog\'{\i}a (CONACYT), Mexico; Ministerie van Onderwijs, Cultuur en Wetenschap, Nederlandse Organisatie voor Wetenschappelijk Onderzoek (NWO), Stichting voor Fundamenteel Onderzoek der Materie (FOM), Netherlands; National Centre for Research and Development, Grant Nos.ERA-NET-ASPERA/01/11 and ERA-NET-ASPERA/02/11, National Science Centre, Grant Nos. 2013/08/M/ST9/00322 and 2013/08/M/ST9/00728, Poland; Portuguese national funds and FEDER funds within COMPETE - Programa Operacional Factores de Competitividade through Funda\c{c}\~{a}o para a Ci\^{e}ncia e a Tecnologia, Portugal; Romanian Authority for Scientific Research ANCS, CNDI-UEFISCDI partnership projects nr.20/2012 and nr.194/2012, project nr.1/ASPERA2/2012 ERA-NET, PN-II-RU-PD-2011-3-0145-17, and PN-II-RU-PD-2011-3-0062, the Minister of National  Education, Programme for research - Space Technology and Advanced Research - STAR, project number 83/2013, Romania; Slovenian Research Agency, Slovenia; Comunidad de Madrid, FEDER funds, Ministerio de Educaci\'{o}n y Ciencia, Xunta de Galicia, European Community 7th Framework Program, Grant No. FP7-PEOPLE-2012-IEF-328826, Spain; The Leverhulme Foundation, Science and Technology Facilities Council, United Kingdom; Department of Energy, Contract No. DE-AC02-07CH11359, DE-FR02-04ER41300, and DE-FG02-99ER41107, National Science Foundation, Grant No. 0450696, The Grainger Foundation, USA; NAFOSTED, Vietnam; Marie Curie-IRSES/EPLANET, European Particle Physics Latin American Network, European Union 7th Framework Program, Grant No. PIRSES-2009-GA-246806; and UNESCO.

\end{document}